\documentclass[journal,10pt]{IEEEtran}

\usepackage{graphicx}
\usepackage{algorithm}
\usepackage{algorithmic}

\usepackage{booktabs}

\usepackage{subfigure}
\usepackage{multicol}
\usepackage{multirow}
\usepackage{amsmath}
\usepackage{bm}
\usepackage{cite}
\usepackage{gensymb}
\usepackage{slashbox}
\usepackage{amsfonts}
\usepackage{mathrsfs}
\usepackage{amsmath}
\usepackage{color}
\usepackage{balance}
\usepackage{bm}
\usepackage{setspace}
\usepackage{url}
\usepackage{setspace}

\usepackage{array}
\newcommand{\PreserveBackslash}[1]{\let\temp=\\#1\let\\=\temp}
\newcolumntype{C}[1]{>{\PreserveBackslash\centering}p{#1}}
\usepackage[flushleft]{threeparttable}

\begin{document}

\bibliographystyle{IEEEtran} 

\title{Eliminating NB-IoT Interference to LTE System: a Sparse Machine Learning Based Approach}

\author{Sicong Liu, \emph{Member, IEEE}, Liang Xiao, \emph{Senior Member, IEEE}, \\
Zhu Han, \emph{Fellow, IEEE}, and Yuliang Tang, \emph{Member, IEEE}
\thanks{

This work was supported by the National Natural Science Foundation of China (Grant No. 61731012, 91638204, 61671396, 61371081, and 61871339), the Natural Science Foundation of Fujian Province of China (Grant No. 2019J05001), the Fundamental Research Funds for the Central Universities of China, US MURI AFOSR MURI 18RT0073, NSF CNS-1717454, CNS-1731424, CNS-1702850, CNS-1646607, and the open research fund of the National Mobile Communications Research Laboratory, Southeast University (No.2018D08). (Corresponding Author: Sicong Liu.)

Sicong Liu, Liang Xiao, and Yuliang Tang are with Department of Communication Engineering \& Key Laboratory of Digital Fujian on IoT Communication, Architecture and Security Technology, Xiamen University, Xiamen 361005, P. R. China.
 (E-mail: \{liusc, lxiao, tyl\}@xmu.edu.cn)

 Zhu Han is with  the University of Houston, Houston, TX 77004 USA (e-mail:zhan2@uh.edu), and also with the Department of Computer Science and Engineering, Kyung Hee University, Seoul, South Korea, 446-701.
 }

}
\maketitle
\begin{abstract}
Narrowband internet-of-things (NB-IoT) is a competitive 5G technology for massive machine-type communication scenarios, but meanwhile introduces narrowband interference (NBI) to existing broadband transmission such as the long term evolution (LTE) systems in  enhanced mobile broadband (eMBB) scenarios. In order to facilitate the  harmonic and fair coexistence in wireless heterogeneous networks, it is important to eliminate NB-IoT interference to LTE systems.  In this paper, a novel sparse machine learning based framework and a sparse combinatorial optimization problem is formulated for accurate NBI recovery, which can be efficiently solved using the proposed iterative sparse learning algorithm called sparse cross-entropy minimization (SCEM). To further improve the recovery accuracy and convergence rate, regularization is introduced to the loss function in the enhanced algorithm called regularized SCEM. Moreover, exploiting the spatial correlation of NBI, the framework is extended to multiple-input multiple-output systems. Simulation results demonstrate that the proposed methods are effective in eliminating NB-IoT interference to LTE systems, and significantly outperform the state-of-the-art methods.

\end{abstract}


\IEEEpeerreviewmaketitle

\section{Introduction}
\IEEEPARstart{W}{ITH} the rapid development of the upcoming technologies of 5G new radio, the extensive research on enhanced mobile broadband (eMBB), massive machine-type communications (mMTC), and ultra-reliable low latency communications (URLLC) has drawn dramatically increasing attention from both academia and industry~\cite{ZHan17WCM,Drira15,ZHan15TWC}. To satisfy the prospects of 5G, not only tremendous improvements of the aforementioned new radio techniques need to be achieved, but also the harmonic and fair coexistence of heterogeneous networks and the compatibility between 4G and 5G systems should be taken great care of~\cite{YLi15JSAC}. Due to the scarcity of the spectrum suitable for wireless electromagnetic transmission, many various existing and emerging communication systems are deployed close to each other, or even overlapping in spectrum, which inevitably results in intensive interference~\cite{LXiao18TVT}. As a typical example, the narrowband internet-of-things (NB-IoT) system is deployed reusing the spectrum of long term evolution (LTE), occupying the spectrum of LTE when operating in the ``in-band'' mode~\cite{YSun_IoTJ_18, YanjunLi_IoTJ_18, NB-IoT1}. NB-IoT is a promising and emerging technology to support the prospect of mMTC in 5G new radio, capable of interconnecting a large amount of nodes with very low power consumption and narrow bandwidth~\cite{YLi_IoTJ_18,Tsoukaneri_IoTJ, NB-IoT2}. Since LTE and LTE-Advanced (LTE-A) with the cyclic-prefixed orthogonal frequency division multiplexing (CP-OFDM) modulation are dominating technologies in 4G era~\cite{LTE1, LTE2, Ghosh2010LTE}, the interference from NB-IoT systems should be properly tackled so that the smooth transition from 4G to 5G can be done~\cite{NB-IoT3, LPQian_IoTJ_18}. In the process of the deployment of 5G eMBB facilities, it is also important to mitigate the interference from NB-IoT if the utilized spectrum is overlapping.

However, how to mitigate or eliminate the interference between NB-IoT and LTE systems still remains an open issue, which has not been sufficiently investigated in literature yet. Since the bandwidth of NB-IoT is sufficiently small compared with that of LTE, the interference from NB-IoT can be regarded as a certain kind of narrowband interference (NBI). Although there are plenty of conventional methods to combat against NBI in literature~\cite{S-Kai,R-Nilsson,D-Darsena,Darsena08,Coulson06}, useful data might be lost using the conventional methods, or the information of statistics or locations of the NBI should be priorly known, or a large amount of virtual sub-carriers were consumed, which limited the efficiency and applicability of the conventional methods.

Recently, emerging sparse recovery methods are introduced to NBI estimation, exploiting the sparsity property of NBI, especially the compressed sensing (CS) theory based methods are drawing great attention~\cite{D-Donoho}. Nevertheless, the state-of-the-art CS-based methods are mostly designed for non-CP-OFDM systems, or the estimation is carried out at the preamble, which might turn out inaccurate for the payload data frames. Besides, it is difficult to design a practical observation matrix with satisfactory restricted isometry property (RIP) required by CS-based methods~\cite{D-Donoho}. Thus, the performance is limited when the conditions of background noise or sparsity level are unideal. Sparse Bayesian learning (SBL), as another sparse recovery theory, was proposed~\cite{Z-Zha} to solve block sparse recovery problems, but prior information of the block partition and the statistics of the unknown signal were required, and the stringent parametric assumptions of the NBI were impractical.

Different from the aforementioned existing schemes, the emerging and powerful machine learning theory and techniques, drawing tremendous research attention recently, can be a great inspiration to achieve a both efficient and reliable method of NBI recovery. In the research on machine learning, cross-entropy (CE) has been exploited as the loss function to train deep neural networks~\cite{kroese2013cross}. Nevertheless, the conventional CE method was not designed for sparse approximation. Moreover, the state-of-the-art research on sparse machine learning  based NBI recovery using iterative cross-entropy guided training is insufficient in literature. To fill this gap,  a sparse machine learning inspired probabilistic framework is formulated, and a novel algorithm called sparse CE minimization  (SCEM) is proposed to iteratively learn the support distribution. The proposed method is capable of learning and recovering the NBI more efficiently  and  more accurately than  state-of-the-art counterparts, supporting the harmonic coexistence of NB-IoT and LTE systems.

\color{black}
The main contributions are listed as follows:

\begin{itemize}
\item The theory of sparse machine learning  with the method of CE-guided training  is  introduced to the area of NBI recovery for the first time. A novel probabilistic framework of sparse machine learning is formulated to recover and eliminate the NB-IoT interference to the LTE system, with higher spectral efficiency and recovery accuracy than the existing methods.

\item A novel algorithm called SCEM based on sparse machine learning is proposed for NBI recovery, which iteratively learns the NBI support distribution guided by the CE as the loss function. An enhanced algorithm called regularized SCEM (RSCEM) is proposed by regularizing the loss function, which achieves better recovery accuracy and convergence rate.

\item The proposed framework is extended to MIMO systems to utilize the spatial correlation of the NBI at multi-antennas. Thus the simultaneous SCEM (S-SCEM) algorithm is formulated, which combines the contributions from multiple antennas and simultaneously recovers the common support of the NBI to further improve the spectral efficiency and accuracy.
\end{itemize}

 The rest of this paper is organized as follows: The related works are presented in Section II. The system model is presented in Section III. The main contribution of this paper, the proposed probabilistic framework formulation and the proposed algorithms of sparse machine learning for NBI recovery, are described in detail in Section IV. The performance of the proposed algorithms is evaluated through computer simulations in Section V, which is followed by the conclusions in Section VI. \color{black}

$Notation$. Matrices and column vectors are denoted by boldface letters; frequency-domain and time-domain vectors are denoted by boldface vectors with tilde $\tilde{\bf v}$ and without tilde $\bf v$, respectively; $(\cdot)^\dagger$ and $(\cdot)^H$ denote the pseudo-inversion operation and conjugate transpose, respectively; $\Arrowvert \cdot \Arrowvert_r$  represents the $\ell_r$ norm operation; $|\Pi|$ denotes the cardinality of the set $\Pi$; $\mathbf{v} \arrowvert_\Pi$ denotes the entries of the vector $\mathbf{v}$ in the set of $\Pi$; $\mathbf{A}_\Pi$ represents the sub-matrix comprised of the columns of the matrix $\mathbf{A}$ indexed by $\Pi$; $\Pi^c$ denotes the complementary set of $\Pi$; $\textup{supp}(\mathbf{v})$ denotes getting the support of $\mathbf{v}$.

$Synonyms$. BSBL (Block Sparse Bayesian Learning). CE (Cross-Entropy). CP (Cyclic Prefix). CRLB (Cramer-Rao Lower Bound). CS (Compressed Sensing). FTE (Frequency Threshold Excision). IBI (Inter-Block Interference). INR (Interference-to-Noise Ratio). LTE (Long Term Evolution). MIMO (Multiple-Input Multiple-Output). MSE (Mean Square Error). NBI (NarrowBand Interference). NB-IoT (NarrowBand Internet-of-Things). NLL (Negative Logarithm Likelihood). OFDM (Orthogonal Frequency Division Multiplexing). PA-SAMP (Priori Aided Sparsity Adaptive Matching Pursuit). RIP (Restricted Isometry Property). RSCEM (Regularized Sparse Cross-Entropy Minimization). SCEM (Sparse Cross-Entropy Minimization). S-SCEM (Simultaneous Sparse Cross-Entropy Minimization). SAMP (Sparsity Adaptive Matching Pursuit).

\begin{table}[h!]

\caption{Frequently Used Symbols}
\renewcommand{\arraystretch}{1.4}
\vspace{-15pt}
\small
\begin{center}
\begin{tabular}{*{1}{m{0.8cm}}|*{1}{m{2.5cm}}|*{1}{m{0.8cm}}|*{1}{m{2.5cm}}}
 \hline\hline
    \bf symbol & \bf concept & \bf symbol &  \bf concept\\
  \hline
    ${{{\bf{\tilde e}}}_{{\rm{B}}}}$ & block sparse NBI vector &  $\Delta{{{\bf{\tilde e}}}_{{\rm{B}}}}$ &  differential NBI vector\\
  \hline
  $\Delta {\bf{p}}$ & NBI measurement vector &  $\epsilon$ &  AWGN error norm threshold\\
  \hline
  ${\bf{S}}_{G,N}$ &  $G \times N$ selection matrix  & ${{\bf{F}}_N}$ &  $N \times N$ IDFT matrix\\
     \hline
  ${\bf{\Psi }}$ &  observation matrix  & ${{\bf{q}}^{(k)}}$ &  current support distribution\\
     \hline
    ${\Pi _j^{(k)}}$ &  candidate support  & ${\Pi _{[j]}^{(k)}}$ &  favorable support\\
     \hline
  $r_j^{(k)}$ &  residue error norm  & $\bar{r}_j^{(k)}$ & weighted average residue error norm\\
     \hline
   $N_c$ &  candidate supports number  & $N_f$ &  favorable supports number\\
     \hline
   ${\lambda _{[j]}}$ &  regularization weight parameter  &$I_{\textup m}$ &  maximum iteration number\\
     \hline
  $G$ & measurement vector length & $N$ & OFDM sub-carrier number\\
     \hline
  $N_{\textup R}$ & MIMO receive antenna number &$K$ &  NBI sparsity level\\
  \hline \hline
\end{tabular}
\end{center}
\normalsize
\end{table}

\section{Related Works}
{ Some coexistence simulation results for in-band and guard band scenarios between NB-IoT and legacy systems are provided for initial analysis in the 3GPP technical document~\cite{NB-IoT0}, which shows significant interference between NB-IoT and LTE systems.} Ratasuk \emph{et al} provided an analysis of the impacts of the NB-IoT signal on the link budget and block error rate performance of the LTE system~\cite{ratasuk2017analysis}. Kim  \emph{et al} investigated the interference between NB-IoT and LTE systems in the ``in-band'' mode~\cite{kim2017analysis}. Wang and Wu gave an analysis of the coexistence between NB-IoT and LTE for the stand-alone mode, and studied the effects of NB-IoT to the performance of uplink LTE transmission~\cite{wang2016coexistence}.

Since the problem of the coexistence between NB-IoT and LTE systems is vital, there have been some conventional methods to combat against NBI. A commonly adopted approach is to directly null out the sub-carriers where NBI is present, called frequency threshold excision (FTE)~\cite{S-Kai}. Nilsson proposed a linear minimum mean square error based method to estimate NBI~\cite{R-Nilsson}. A successive interference cancelation approach mitigating the NBI in a recursive manner was introduced in~\cite{D-Darsena}. A soft decision based successive NBI cancellation method was further proposed by Darsena \emph{et al} in~\cite{Darsena08}. Coulson designed a time-domain notch filter for NBI suppression based on linear prediction criterion before discrete Fourier transform at the transmitter~\cite{Coulson06}. The limitation of conventional methods mainly lies in that useful data might be lost, and that the statistics information or plenty of virtual sub-carriers are required.

To overcome the limitations of conventional methods, the CS theory, as a newly emerged powerful approach for sparse recovery, can be utilized to deal with the NBI estimation problem.
CS-based methods were first investigated by Al-Dhahir \emph{et al}, utilizing the null space to obtain the measurements of NBI for OFDM systems~\cite{A-Gom2,Dhahir4}. In this work, the NBI could be recovered by using CS-based greedy algorithms.
There have been studies on different CS-based greedy algorithms, such as subspace pursuit (SP)~\cite{W-Dai} proposed by W. Dai \emph{et al} and sparsity adaptive matching pursuit (SAMP)~\cite{T-Do} proposed by T. Do \emph{et al}. The SP algorithm is able to recover sparse signals with or without noise disturbance costing low complexity~\cite{W-Dai}.
The SAMP algorithm is designed to be adaptive to variant sparsity levels of the NBI. By dividing the iteration process into multiple stages, the SAMP algorithm is able to recover the sparse signal by iterative matching pursuit of the support basis without knowing its sparsity level~\cite{T-Do}.

Other CS-based methods were proposed to estimate the NBI, exploiting the time-domain training guard interval of time-domain synchronous OFDM (TDS-OFDM) systems~\cite{S-Liu5} or the preamble in the frame header~\cite{S-Liu6}. In the work of~\cite{S-Liu5}, the algorithm of priori aided SAMP (PA-SAMP) was proposed as an improvement of the classical algorithm SAMP~\cite{T-Do}, which makes use of the prior information of the partial NBI support acquired by the coarse power threshold method. Then the prior information was exploited in the initialization and iteration process to reduce the complexity and improve the accuracy. The two-dimensional correlation of the NBI was exploited in the framework of multiple measurements and structured CS, in literature~\cite{S-Liu6}. The two-dimensional measurement data were obtained from the preambles in multiple receive antennas, and then utilized for the structured CS based recovery of the NBI.
Another sparse recovery theory, sparse Bayesian learning (SBL), was proposed in~\cite{Z-Zha} and has been utilized to effectively estimate the impulsive noise~\cite{JLinJSAC}. A block SBL (BSBL) based method of estimating the NBI generated by NB-IoT was proposed in~\cite{SLiu17TCOM}, which is an improvement of the SBL-based method in~\cite{Z-Zha}. The BSBL-based method employed parametric Bayesian inference iteratively to estimate the unknown deterministic parameters of the block sparse NBI~\cite{SLiu17TCOM}. However, the major limitation of CS-based methods is that the CS theory requires that an observation matrix with satisfactory RIP should be designed~\cite{D-Donoho}, which is difficult in practice. Furthermore, the performance is limited when the intensity of the background noise or sparsity level is large.

Machine learning has become a popular research trend in recent years, with many applications in the area of sparse composite regularization~\cite{GGui18Co}, anti-jamming~\cite{HZhang18,LXiao18Two}, as well as wireless communications~\cite{GGui18deep}. A reinforcement learning based scheme was proposed in literature~\cite{HZhang18} for ultra-dense networks, which adaptively learns the policy of power control to improve the efficiency while mitigating the inter-cell interference. A two-dimensional anti-jamming mobile communication scheme based on reinforcement learning was proposed in literature~\cite{LXiao18Two}, where a mobile
device can achieve an optimal communication policy without the need to know the jamming and interference model in a dynamic game framework. As an important method in machine learning, the CE method is usually utilized for training deep neural networks and machine learning models, which has well solved many learning tasks such as pattern recognition, object classification and so on~\cite{hinton2006reducing,Robert2012Machine}. Recently, a machine learning based method exploiting CE was proposed in~\cite{GGui19improved} to improve hybrid precoding performance for mmWave massive MIMO systems, which introduced it to wireless communications research. Previously, the CE method was also adopted to solve combinatorial optimization problems in literature, which outperforms the brute-force approach~\cite{kroese2013cross, de2005tutorial}. Different from the state-of-the-art methods, the proposed solution in this work introduces sparse machine learning to NBI estimation, and a novel algorithm based on CE minimization is proposed to efficiently learn the NBI support, which improves both the spectral efficiency and the estimation accuracy compared with existing approaches.

\color{black}

\section{System Model}
\subsection{Signal Model of LTE}
\begin{figure}[t]
\centering
\includegraphics[width=0.42\textwidth]{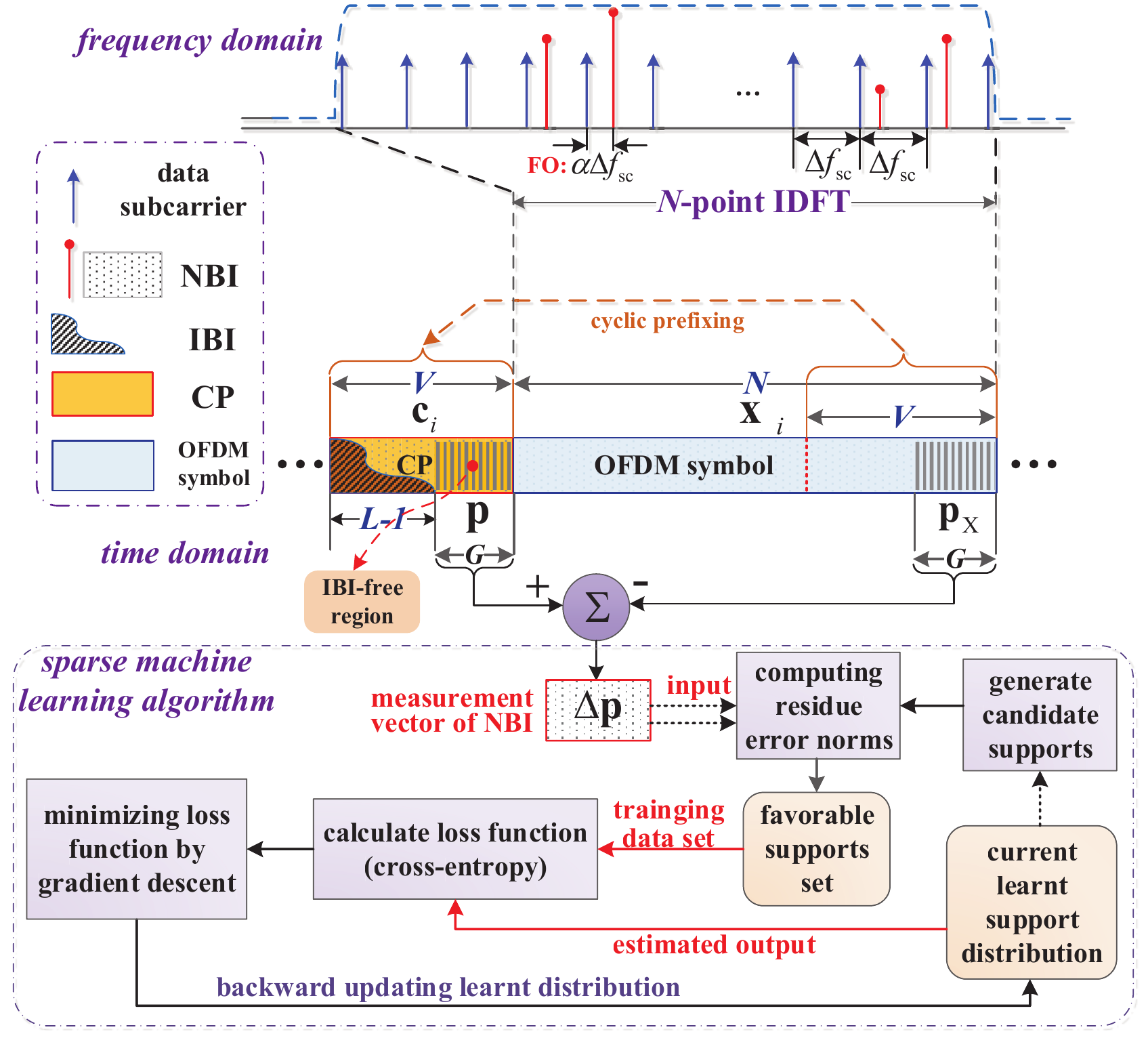}
\caption{Temporal Differential Measuring of NBI from CP-OFDM Frames and Sparse Machine Learning Based Framework Formulation for NBI Recovery in LTE Systems.}
\label{BS_TDM}
\end{figure}

As adopted in 3GPP standards of LTE~\cite{LTE1, LTE2}, the CP-OFDM frame structure is composed of  the length-$N$ OFDM block, where $N$ is the number of sub-carriers with the sub-carrier spacing of $\Delta f_{\textup{sc}}$, and the \mbox{length-$V$} CP in front, which is formed by the last $V$ samples of the OFDM block, as illustrated in Fig.~\ref{BS_TDM}.

 After transmitted in the wireless multi-path fading channel with the channel impulse response (CIR) ${\bf{h}}_i = {\left[ {{h_{i,0}},{h_{i,1}}, \cdots ,{h_{i,L - 1}}} \right]^T}$ in the presence of the NBI generated by the NB-IoT signal, the received $i$-th CP $\mathbf{c}_i = {\left[ {{c_{i,0}},{c_{i,1}}, \cdots ,{c_{i,V - 1}}} \right]^T}$  before the $i$-th  OFDM block $\mathbf{x}_{i}$ is represented as \color{black}
\begin{equation}\label{p_i}
\mathbf{c}_i = \mathbf{\Psi}_{\textup{CP}}\mathbf{h}_i + {\mathbf{e}}_i + \mathbf{w}_i,
\end{equation}
where ${\mathbf{e}}_i = {\left[ {e_{i,0},e_{i,1}, \cdots ,e_{i,V - 1}} \right]^T}$ denotes the time-domain NBI vector located at the CP, ${{\bf{w}}_i}$ denotes the additive white Gaussian noise (AWGN) vector with zero mean and variance of $\sigma_w^2$, and ${{\bf{\Psi }}}_{\textup{CP}}{\bf{h}}_i$ denotes the received CP, with the matrix ${{\bf{\Psi }}}_{\textup{CP}} \in \mathbb{C}^{V\times L}$ represented as
{\setlength\arraycolsep{4pt}
\begin{equation*}\label{Equ:Phi}
{\left[
\begin{array}{*{20}{c}}
{{x_{i,N-V}}}&{{x_{i-1,N-1}}}&{ {x_{i-1,N-2}}} &{ \cdots    }&{ {x_{i-1,N-L+1}}}\\
{{x_{i,N-V+1}}}&{x_{i,N-V}}&{ {x_{i-1,N-1}}}  &{ \cdots    }&{ {x_{i-1,N-L+2}}} \\
{{x_{i,N-V+2}}}&{x_{i,N-V+1}}&{ x_{i,N-V}   }&{ \cdots    }&{ {x_{i-1,N-L+3}}} \\
 \vdots & \vdots & \vdots & \ddots & \vdots \\
 {{x_{i,N-V+L-2}}}&{x_{i,N-V+L-3}}&{ x_{i,N-V+L-4}}&{ \cdots    }&{ x_{i-1,N-1} }\\
 {{x_{i,N-V+L-1}}}&{x_{i,N-V+L-2}}&{ x_{i,N-V+L-3}}&{ \cdots    }&{ x_{i,N-V} }\\
 {{x_{i,N-V+L}}}&{x_{i,N-V+L-1}}&{ x_{i,N-V+L-2}}&{ \cdots    }&{ x_{i,N-V+1} }\\
 \vdots & \vdots & \vdots & \ddots & \vdots \\
  {{x_{i,N-1}}}&{{x_{i,N-2}}}&{{x_{i,N-3}}}&{ \cdots   }&{{x_{i,N-L}}}
\end{array} \right]}
\end{equation*}}

\noindent  The entries $\{ {x_{i - 1,n}}\} _{n = N - L + 1}^{N - 1}$ in the matrix ${{\bf{\Psi }}}_{\textup{CP}}$ above represent the last $L-1$  samples of the preceding $(i-1)$-th OFDM block $\mathbf{x}_{i-1}$, which causes inter-block-interference (IBI) on the following $i$-th CP. \color{black} Since $\mathbf{x}_{i-1}$ only causes IBI on the first $L-1$ samples of the $i$-th CP as illustrated in Fig.~\ref{BS_TDM}, the last $G=V-L+1$ samples of $\mathbf{c}_i$ will form the \emph{IBI-free region} given by
\begin{equation}\label{IBI}
{\bf{p}}_i = {\left[ {{c_{i,L-1}},{c_{i,L}}, \cdots ,{c_{i,V - 1}}} \right]^T} = {\bf{S}}_{G,V}\mathbf{c}_i,
\end{equation}
\noindent where  ${\bf{S}}_{G,V}$ denotes the selection matrix composed of the last $G$ rows of the $V \times V$ identity matrix ${\bf{I}}_V$. The IBI-free region exists in practical broadband transmission systems because a common rule for system design is to configure the guard interval length $V$ to be much larger than the maximum channel delay spread $L$ in the worst case to avoid IBI between OFDM symbols, which is specified in standards and supported in literature~\cite{Ghn, LTE1, SLiu17TCOM}.

For simplicity of notations, the subscript of $i$ denoting the frame number is omitted  in the following content of this paper when there is no ambiguity about the current frame number, unless otherwise clearly stated.
Then the IBI-free region can be rewritten as
\begin{equation}\label{q_i}
{\bf{p}} = {\bf{S}}_{G,V}{{\bf{\Psi }}_{\textup{CP}}}{\bf{h}} + {\mathbf{e}}+ {\bf{w}},
\end{equation}
where ${\bf{p}}$, ${\mathbf{e}}$, and ${\bf{w}}$ consist of the last $G$ entries of ${\bf{c}}_i$, ${\mathbf{e}}_i$, and ${\bf{w}}_i$ in \eqref{p_i}, respectively, while ${\bf{S}}_{G,V}{{\bf{\Psi }}_{\textup{CP}}} \in \mathbb{C}^{G \times L}$ is composed of the last $G$ rows of ${{\bf{\Psi }}_{\textup{CP}}}$ without the IBI component. Since the CP is the same with the last $V$ samples of the OFDM block, there is a duplicate of the IBI-free region ${\bf{p}}$ at the last $G$ samples of its subsequent OFDM block, which can be denoted by ${\bf{p}}_{\textup{X}}$ given by
\begin{equation}\label{qX_i}
{\bf{p}}_{\textup{X}} = {\bf{S}}_{G,V}{{\bf{\Psi }}_{\textup{CP}}}{\bf{h}} + {\mathbf{e}}_{\textup{X}} + {\bf{w}}_{\textup{X}},
\end{equation}
where ${\mathbf{e}}_{\textup{X}}$ and ${\bf{w}}_{\textup{X}}$ denote the length-$G$ time-domain NBI and AWGN vectors at the end of the OFDM block, respectively.

\subsection{{NBI Model Generated by NB-IoT}}\label{III-B}
In LTE systems, the NB-IoT signal working in the ``in-band'' mode at the spectrum of LTE generates NBI to the receivers of the LTE system~\cite{JXu_IoTJ18}.   The widely adopted model of the NBI  in the frequency domain is the superposition of several tone interferers, and each tone interferer is modeled by a band-limited Gaussian noise (BLGN) with the power spectral density (PSD) of $N_{0,\textup{NBI}} = \sigma^2_e$~\cite{D-Ume1}.  \color{black} The  frequency-domain location of the tone interferers can be randomly distributed among all $N$ sub-carriers~\cite{A-Ton2, D-Ume1}, and different tone interferers are mutually independent~\cite{A-Ton2}. Let ${{{\bf{\tilde e}}}_i}=[{{\tilde e}_{i,0}},{{\tilde e}_{i,1}},\cdots,{{\tilde e}_{i,N-1}}]^T$ denote the frequency-domain NBI vector associated with the CP, and then each entry of the corresponding time-domain NBI signal ${\mathbf{e}}_i$  can be represented as
\begin{equation}\label{e_ik}
e_{i,n} = \sum_{k \in \Pi}{\tilde e_{i,k}\cdot\exp ( \frac{j2\pi kn}{N})}, ~~n=0,1,\cdots,V-1,
\end{equation}
\noindent where ${\Pi = \left\{ {k\left| {\tilde e_{i,k} \ne 0} \right., k = 0,1, \cdots ,N - 1} \right\}}$ is the set of the indices of nonzero entries, which is defined as the \emph{support}. The sparsity level $K$ is defined by the number of nonzero entries, which is much smaller than the signal dimension, i.e., ${K = |\Pi|\ll N}$.    The interference-to-noise ratio (INR) $\gamma$  is used to represent the intensity of the NBI, defined by $\gamma = {\mathbb{E}\{{{\mathscr{P}_e}}\} \mathord{\left/ {\vphantom {{{P_e}} {\sigma _{w}^2}}} \right. \kern-\nulldelimiterspace} {\sigma_w^2}}$, where ${\mathscr{P}_e} = \sum\nolimits_{k \in \Pi }^{} {|\tilde e_{i,k}{|^2}} /K$  denotes the average power. \color{black} Since the tone interferers are BLGN as described, the average power is $\mathbb{E}\{{{\mathscr{P}_e}}\} = \sigma^2_e$, yielding the INR $\gamma= \sigma^2_e/\sigma_w^2$.

Since the bandwidth of NB-IoT is sufficiently small compared with that of LTE~\cite{JChen_IoTJ_17}, the NBI generated by NB-IoT can be modeled as a sparse vector in the frequency domain, which has only few nonzero entries compared with the number of sub-carriers. The nonzero entries of the NBI are not necessarily located exactly at the frequencies of the OFDM sub-carriers in LTE, because in practice there might be a fractional frequency offset (FO) for the NB-IoT working frequency with respect to the OFDM sub-carriers. Thus, the generalized NBI model will become a block sparse vector due to the spectral leakage~\cite{Sohail2012}. Then the frequency-domain block sparse NBI vector ${{{\bf{\tilde e}}}_{{\rm{B}}}}=[{{\tilde e}_{{\rm{B}},0}},{{\tilde e}_{{\rm{B}},1}},\cdots,{{\tilde e}_{{\rm{B}},N-1}}]^T$ associated with the CP can be represented as
 \begin{equation}\label{e_Bi}
{{{\bf{\tilde e}}}_{{\rm{B}}}} =  {{\bf{F}}_N^H{{\bf{\Lambda }}_{{\rm{FO}}}}{{\bf{F}}_N}}{{{\bf{\tilde e}}}_i},
\end{equation}
\noindent where ${\bf{F}}_N$ denotes the $N \times N$ inverse discrete Fourier transform (IDFT) matrix with the entry $\{{\bf{F}}_N\}_{m,n} = \exp (j2\pi mn/N)/\sqrt{N}$, and \small ${{\bf{\Lambda }}_{{\rm{FO}}}}=\textup{diag}\{1,\exp(j2\pi\alpha/N),\cdots,\exp(j2\pi\alpha(N-1)/N)\}$ \normalsize is the FO matrix, whose value of offset frequency can be modeled by a uniformly distributed variable $\alpha \in \textup{U}(-1/2,1/2]$~\cite{Sohail2012}. Transforming the frequency-domain NBI signal~\eqref{e_Bi} to the time domain by partial IDFT, the NBI vector associated with the IBI-free region in \eqref{q_i} is obtained as
\begin{equation}\label{e_i1}
{\mathbf{e}} = {{\bf{S}}_{G,N}}{{\bf{F}}_N}{{{\bf{\tilde e}}}_{{\rm{B}}}}.
\end{equation}

There is a useful feature of NBI called \emph{temporal correlation}, which can be utilized for measuring the NBI from the compound received signal containing both the NBI and the data components. The temporal correlation claims that, the NBI signal usually has invariant support and amplitude over one received OFDM frame of interest. This is because according to experiments and observations, the coherence time of the NBI signal is normally much larger than that of one OFDM symbol, and the working band of the NBI source such as NB-IoT is not changing so fast~\cite{PLC-book, J-Zhang1, SLiu17TCOM}. It is observed that usually the NB-IoT signal working in-band in LTE spectrum is located fixed in certain frequency locations~\cite{NB-IoT1, NB-IoT2}. Temporal correlation is also verified by substantial field tests and experimental observations in real house and apartments~\cite{cortes2010analysis}.

Because of the temporal correlation, the frequency-domain NBI vectors associated with the CP part and the following OFDM block part share the same support and amplitude, with only a phase shift in between:  Let ${{{\bf{\tilde e}}}_{{\rm{B}}\textup{X}}}=[{{\tilde e}_{{\rm{BX}},0}},{{\tilde e}_{{\rm{BX}},1}},\cdots,{{\tilde e}_{{\rm{BX}},N-1}}]^T$ denote the frequency-domain NBI vector  associated with the CP's duplicate in the OFDM block given by~\eqref{qX_i}, where the time-domain representation of ${{{\bf{\tilde e}}}_{{\rm{B}}\textup{X}}}$ is given by
\begin{equation}\label{e_Xi1}
{\mathbf{e}}_{\textup{X}} =  {\bf{S}}_{G,N}{{\bf{F}}_N}{{{\bf{\tilde e}}}_{{\rm{B}}\textup{X}}}.
\end{equation}
\noindent Hence, ${{{\bf{\tilde e}}}_{{\rm{B}}\textup{X}}}$ can be derived by the phase shift of ${{{\bf{\tilde e}}}_{{\rm{B}}}}$ associated with the CP in~\eqref{q_i}, which can be represented as
\begin{equation}\label{e_BXik}
{{\tilde e}_{{\rm{BX}},k}} = {{\tilde e}_{{\rm{B}},k}}\exp \left( {\frac{{j2\pi (k + \alpha )\Delta l_{\rm{B}}}}{N}} \right),k = 0,1, \cdots ,N - 1,
\end{equation}
\noindent where the value of FO $\alpha$ determines the phase to shift, and $\Delta l_{\rm{B}}$ is the corresponding time-domain distance between the CP and its duplicate in the OFDM block.

Note that $\Delta l_{\rm{B}}=N$ as illustrated in Fig.~\ref{BS_TDM}, so it can be further derived that ${{\tilde e}_{{\rm{BX}},k}} = {{\tilde e}_{{\rm{B}},k}}\exp \left( {j2\pi \alpha } \right)$, which yields a simpler relation only related with $\alpha$ given by
\begin{equation}\label{e_BXi}
{{{\bf{\tilde e}}}_{{\rm{B}}\textup{X}}}=\exp \left( {j2\pi \alpha } \right){{{\bf{\tilde e}}}_{{\rm{B}}}}.
\end{equation}

\section{Probabilistic Sparse Machine Learning Based Framework Formulation and Algorithms for NBI Recovery}\label{IV}

In this section, the probabilistic framework of sparse machine learning as well as the sparse combinatorial optimization problem for NBI recovery is firstly formulated in Section~\ref{IV}-A.  Then the proposed sparse machine learning based iterative algorithm called SCEM is introduced in detail in Section~\ref{IV}-B, followed by the enhanced algorithm of RSCEM imposing regularization on the loss function in Section~\ref{IV}-C. Afterwards, the extension of the proposed method to MIMO systems is presented in Section~\ref{IV}-D. \color{black}

\subsection{Probabilistic Sparse Machine Learning Framework Formulation for NBI Recovery}
The ultimate goal of this work is to accurately recover the NBI vector ${{{\bf{\tilde e}}}_{{\rm{B}}\textup{X}}}$ located at the OFDM data block and eliminate it from the data, which can be done by estimating ${{{\bf{\tilde e}}}_{{\rm{B}}}}$ and using the relation in~\eqref{e_BXi}.  Hence, firstly the measurement of the NBI  ${{{\bf{\tilde e}}}_{{\rm{B}}}}$ should be obtained, and a probabilistic sparse machine learning based framework can be formulated to efficiently recover the NBI using the proposed algorithms. \color{black}

The measurement vector of the NBI  can be obtained using the temporal differential measuring operation~\cite{SLiu17TCOM}. Specifically, as illustrated in Fig.~\ref{BS_TDM}, the measurement vector can be obtained by the differential operation between the received IBI-free region ${\bf{p}}$ in \eqref{q_i} and its duplicate ${\bf{p}}_{\textup{X}}$ in \eqref{qX_i} at the end of the OFDM block, which nulls out the cyclic data component ${\bf{S}}_{G,V}{{\bf{\Psi }}_{\textup{CP}}}{\bf{h}}$, yielding the measurement vector  of the NBI
\begin{equation}\label{BSBL1}
\Delta {\bf{p}} = \Delta {\bf{e}} + \Delta {{\bf{w}}},
\end{equation}
where $\Delta {\bf{e}}={\bf{e}}-{\mathbf{e}}_{\textup{X}}$ and $\Delta {{\bf{w}}}= {{\bf{w}}} - {{\bf{w}}}_{\textup{X}}$. Thus by substituting~\eqref{e_i1} and~\eqref{e_Xi1} into~\eqref{BSBL1}, the measurement vector can be rewritten as
\begin{equation}\label{BSBL}
\Delta {\bf{p}} = {\bf{S}}_{G,N}{{\bf{F}}_N}\Delta{{{\bf{\tilde e}}}_{{\rm{B}}}} + \Delta{{\bf{w}}},
\end{equation}
where ${\Delta{{\bf{\tilde e}}_{\rm{B}}}}$ is given by
\begin{equation}\label{de_Bi}
\Delta{{{\bf{\tilde e}}}_{{\rm{B}}}}={{{\bf{\tilde e}}}_{{\rm{B}}}}-{{{\bf{\tilde e}}}_{{\rm{B}}\textup{X}}}=(1-\exp \left( {j2\pi \alpha } \right)){{{\bf{\tilde e}}}_{{\rm{B}}}},
\end{equation}
whose support is the same with that of ${{{\bf{\tilde e}}}_{{\rm{B}}}}$ and ${{{\bf{\tilde e}}}_{{\rm{B}}\textup{X}}}$.

 After obtaining the measurement of the NBI in~\eqref{BSBL}, the probabilistic sparse machine learning framework of NBI recovery can be formulated, by which the support distribution of the NBI can be learnt  using the proposed algorithms.
Because of the sparsity of the frequency-domain NBI vector, it is crucial to recover its support, i.e., the set of the indices of the nonzero entries.   Since the sparsity level of the NBI is $K$, it is required that the unknown NBI vector $\Delta{{{\bf{\tilde e}}}_{{\rm{B}}}} \in \mathbb{C}^N$ to be reconstructed in~\eqref{BSBL} should satisfy \color{black}
\begin{equation}\label{IN_Kmax}
\left\| \Delta{{{\bf{\tilde e}}}_{{\rm{B}}}} \right\|_0 \le K
\end{equation}
\noindent where   $\Arrowvert \cdot \Arrowvert_0$ denotes the $\ell_0$-norm, i.e., the number of nonzero entries. To recover the optimal NBI vector based on the measurement in~\eqref{BSBL}, we should solve the optimization problem given by
\begin{equation}\label{sparse_comb_opt}
\Delta{\hat{\bf{e}}}_{\rm B}^{*} = \mathop{\arg \min}\limits_{\Delta{{\bf{\tilde e}}_{\rm{B}}}} \left\| {\Delta \bf{p}} - {\bf{S}}_{G,N}{{\bf{F}}_N}\Delta{{{\bf{\tilde e}}}_{{\rm{B}}}} \right\|_2, \textup{s.t.} \left\| {\Delta{{\bf{\tilde e}}_{\rm{B}}}} \right\|_0 \le K,
\end{equation}
\noindent where $\Delta{\hat{\bf{e}}}_{\rm B}^{*}$ denotes the optimal NBI vector to be recovered from the measurement ${\Delta \bf{p}}$ in~\eqref{BSBL} that minimizes the residue error norm $r$, with $r$ given by
\begin{equation}\label{residue_norm}
r = \left\| {\Delta \bf{p}} - {\bf{S}}_{G,N}{{\bf{F}}_N}\Delta{{{\bf{\tilde e}}}_{{\rm{B}}}} \right\|_2.
\end{equation}
 In the conventional perspective of signal processing, the problem in~\eqref{sparse_comb_opt} is intractable, because of  the non-convex constraint of $\ell_0$-norm. Since the constraint is a sparse one, it can be regarded as a sparse combinatorial optimization problem. Let $\Xi$ denote the set of all possible supports of sparse vectors satisfying the constraint in~\eqref{IN_Kmax}, we have
 \begin{equation}
\Xi  = \left\{ {\left. \textup{supp}({\Delta{{\bf{\tilde e}}_{\rm{B}}}} \in {\mathbb{C}^N}) \right|{{\left\| {\Delta{{\bf{\tilde e}}_{\rm{B}}}} \right\|}_0} \le {K}} \right\},
\end{equation}
so the size of the set $\Xi$ of possible solutions is given by
\begin{equation}\label{space_cardinality}
\left| \Xi \right| = \sum\limits_{k=0}^{k=K}{\textup{C}_N^{k}} = \sum\limits_{k=0}^{k=K}{\frac{N!}{(N-K)! K!}}.
\end{equation}
It can be noted from~\eqref{space_cardinality} that the possible supports of the solution space is exponentially and combinatorially increasing with the problem size $N$ and $K$.

Some sparse approximation methods, including the popular CS-based theory,  have been exploited to relax the non-convex optimization problem to a tractable one in literature. For instance, the non-convex $\ell_0$-norm constraint in~\eqref{sparse_comb_opt} can be relaxed to the convex $\ell_1$-norm minimization problem~\cite{D-Donoho} as
\begin{equation}\label{l1_minimizaiton}
 \mathop{\arg \min}\limits_{\Delta{{\bf{\tilde e}}_{\rm{B}}}} \left\| {\Delta{{\bf{\tilde e}}_{\rm{B}}}} \right\|_1, \textup{s.t.} \left\| {\Delta \bf{p}} - {\bf{S}}_{G,N}{{\bf{F}}_N}\Delta{{{\bf{\tilde e}}}_{{\rm{B}}}} \right\|_2  \le \epsilon,
\end{equation}
\noindent where $\epsilon$ denotes the error norm bound due to the background AWGN noise $\Delta{{\bf{w}}}$, and thus convex programming can be exploited to solve it~\cite{BP}.
However, the performance of the CS-based methods is dependent on the RIP of the observation matrix~\cite{D-Donoho, han2013compressive}. {Besides, performance degradation could be caused due to intensive background noise and large sparsity level~\cite{D-Donoho}.} The spectral efficiency could still be improved because many measurement samples have to be reserved in the guard interval for CS-based methods~\cite{S-Liu5}.

 To overcome the difficulties of state-of-the-art methods,  a probabilistic sparse machine learning based approach called SCEM is proposed for NBI recovery, which is able to efficiently solve the non-convex sparse combinatorial optimization problem in~\eqref{sparse_comb_opt} without strict prior RIP requirements for the observation matrix ${\bf{S}}_{G,N}{{\bf{F}}_N}$, and much more spectrum-efficient by reducing the cost of measurement data. \color{black} The proposed algorithm significantly develops the conventional CE method~\cite{kroese2013cross} to accommodate the sparse recovery problem, and the unknown sparse NBI signal can be accurately recovered, as described in detail in the next sub-section.

\subsection{Proposed Sparse Machine Learning Inspired Algorithm: Sparse Cross-Entropy Minimization}

\begin{figure}[h]
\centering
\includegraphics[width=0.45\textwidth]{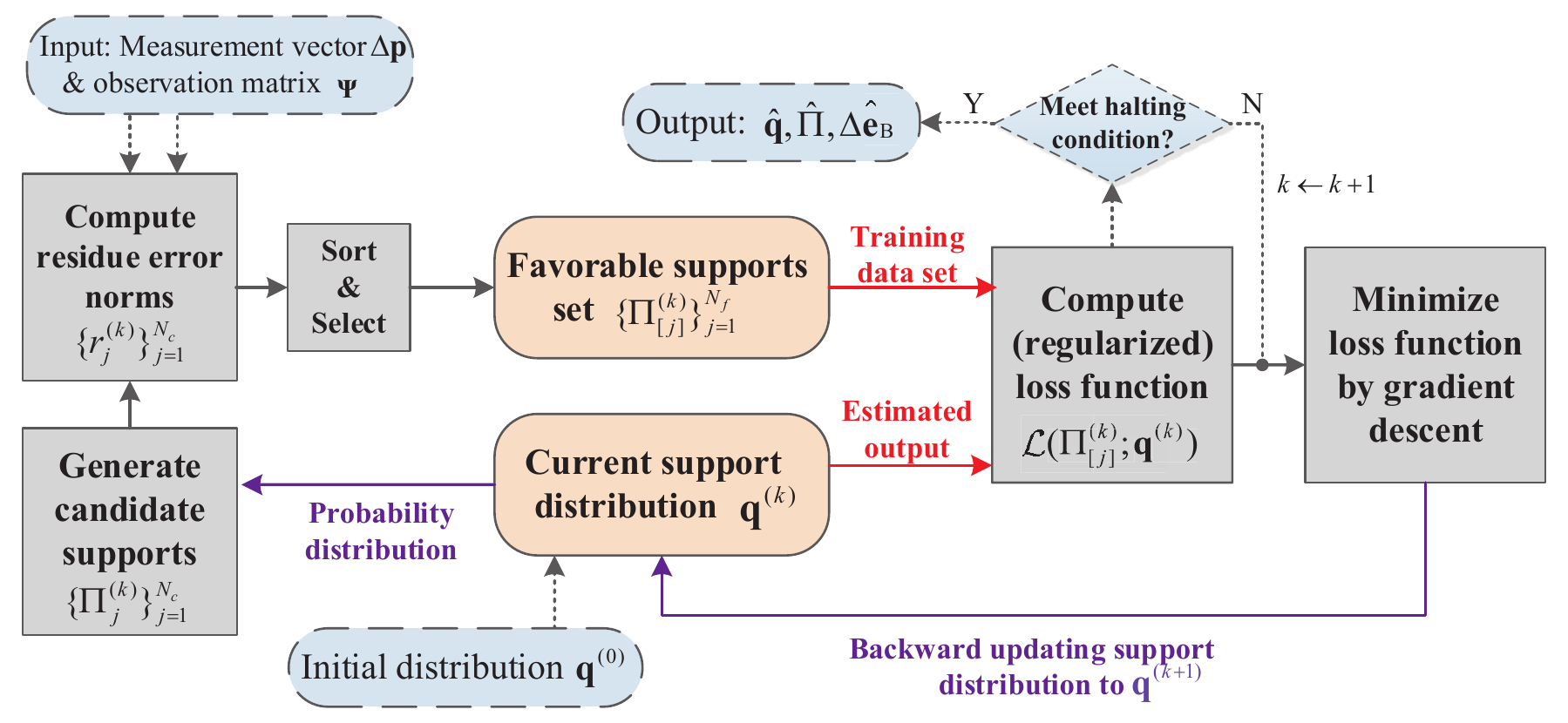}
\caption{Computing flowchart of the Iterative Sparse Machine Learning Based Algorithm of SCEM for NBI Recovery}
\label{SCEM_flowchart}
\end{figure}

Based on the probabilistic framework of sparse learning, the purpose of the SCEM algorithm proposed in this paper is to efficiently solve the sparse combinatorial optimization problem in~\eqref{sparse_comb_opt} by iteratively minimizing the cross-entropy  between the current support distribution and the one minimizing the residue error norm. The pseudo-code of the proposed SCEM algorithm is summarized in \textbf{Algorithm~\ref{Algo1}}, and the computing flowchart of the essential computing modules, parameters, nodes, and data flows of the algorithm is illustrated in Fig.~\ref{SCEM_flowchart}.

It can be observed from Fig.~\ref{SCEM_flowchart} that the proposed sparse machine learning algorithm iteratively learns the probability distribution of the NBI support by minimizing the loss function (i.e., the cross-entropy). In each iteration within the algorithm loop, the algorithm  generates a set of candidate supports randomly based on the current support distribution ${{\bf{q}}^{(k)}}$ (initialized by ${{\bf{q}}^{(0)}}$), and computes the corresponding residue error norms using the measurement vector from the input. After sorting the residue error norms, the set of favorable supports is selected out, which serves as the training data set. Then, the loss function is computed by calculating the cross-entropy between the training data set and the estimated output.  By minimizing the loss function using gradient descent, the support distribution is backward updated to ${{\bf{q}}^{(k+1)}}$ for the next iteration. \color{black}  This process will drive the support distribution gradually to be trained towards the one with minimum estimation error. The iterations continue until the halting condition of the algorithm is met, and the output of the algorithm is thus achieved.

\begin{algorithm}                      
\caption{(\emph{SCEM}): Sparse Cross-Entropy Minimization for Sparse Machine Learning Based NBI Recovery}              
\label{Algo1}                           
\begin{algorithmic}[1]                    
\REQUIRE~~\\
1) Measurement vector ${\Delta \bf{p}}$\\
2) Observation matrix ${\bf{\Psi }} = {\bf{S}}_{G,N}{{\bf{F}}_N}$\\
3) Threshold for residue error norm  $\epsilon$\\
{
4) Candidate supports number $N_c$, favorable supports number $N_f$, maximum iteration number $I_{\textup m}$\\
}
\textbf{\textit{Initialization:}}
\STATE ${{\bf{q}}^{(0)}} \leftarrow \frac{1}{2} \cdot {\overrightarrow {\bf{1}} _{N \times 1}}$  (initial probability distribution of the NBI support)
\STATE $k \leftarrow 0$ (iteration count number)\\

\textbf{\textit{Iterations:}}
\REPEAT
	\STATE Randomly generate $N_c$ \emph{candidate supports} $\{ {\Pi _j^{(k)}} \}_{j = 1}^{{N_c}}$ based on the current support distribution ${\bf{q}}^{(k)}$,
	where each candidate support is generated in a recursive way s.t. $| {\Pi _j^{(k)}} | \le {K}, j = 1, \cdots N_c$\\
	\STATE Compute the corresponding NBI vectors $\{ \Delta{\tilde{\bf{e}}}_{\textup{B},j}^{(k)} \}_{j = 1}^{{N_c}}$, s.t. \\ ${\left. \Delta{\tilde{\bf{e}}}_{\textup{B},j}^{(k)} \right|_{{{\Pi^{(k)}_j}}}} \leftarrow {\bf{\Psi }}_{{{\Pi ^{(k)}_j}}}^\dagger {\Delta \bf{p}}, ~~ {\left. \Delta{\tilde{\bf{e}}}_{\textup{B},j}^{(k)} \right|_{{\Pi ^{(k)c}_j}}}  \leftarrow {\bf{0}}$\\
	\STATE Calculate the corresponding residue error norms $r_j^{(k)} = {\left\| {{\Delta \bf{p}} - {\bf{\Psi }} \Delta{\tilde{\bf{e}}}_{\textup{B},j}^{(k)}} \right\|_2}, j=1,\cdots N_c$\\
	\STATE Sort $\{ {r_j^{(k)}} \}_{j = 1}^{{N_c}}$ in the ascending order as\\ $r_{[1]}^{(k)} \le r_{[2]}^{(k)} \le  \cdots  \le r_{[{N_c}]}^{(k)}$\\
	\STATE Select the $N_f$ smallest residue error norms $\{ {r_{[j]}^{(k)}} \}_{j = 1}^{{N_f}}$, and set the corresponding supports $\{ {\Pi_{[j]}^{(k)}} \}_{j = 1}^{{N_f}}$ as the \emph{favorable supports}\\
	\STATE Update the probability distribution of NBI support to ${\bf{q}}^{(k+1)}$ by minimizing the CE based on~\eqref{q_CEM}\\
	\STATE $k \leftarrow k + 1$\\
\UNTIL  {${r_{[1]}^{(k-1)}} \le \epsilon $ or $k > I_{\textup m}$}  ~~~~~~~  (halting condition)\\
\color{black}

\ENSURE~~\\
1) Learnt support probability distribution ${\hat{\bf{q}}} = {{\bf{q}}^{(k)}}$\\
2) Recovered NBI support ${\hat{\Pi}} = {{ \Pi}^{(k-1)}_{[1]}}$\\
3) Recovered sparse NBI vector $\Delta{\hat{\bf{e}}}_{\rm B} =  \Delta{\tilde{\bf{e}}}_{\rm B,[1]}^{(k-1)}$ \\
\end{algorithmic}
\end{algorithm}

The overall structure and explanations of \textbf{Algorithm~\ref{Algo1}} are described as follows:

Phase 1- Input.  The measurement vector ${\Delta \bf{p}}$, the observation matrix ${\bf{\Psi }}$, the residue error norm threshold $\epsilon$ given in~\eqref{l1_minimizaiton}, and the number of candidate supports and favorable supports, i.e., $N_c$ and  $N_f$, are input to the algorithm. \color{black}

Phase 2 - Initialization. The initial probability distribution of the NBI support is set as ${{\bf{q}}^{(0)}} \leftarrow \frac{1}{2} \cdot {\overrightarrow {\bf{1}} _{N \times 1}}$, where ${{\bf{q}}^{(k)}} = {[q_0^{(k)},q_1^{(k)} \cdots q_{N - 1}^{(k)}]^T} $, and $q_{n}^{(k)}$ denotes the probability that the $n$-th entry is in the NBI support ${\Pi ^{(k)}}$, i.e.,
\begin{equation}
\Pr (n \in {\Pi ^{(k)}}) = q_n^{(k)},n = 0, \cdots N - 1.
\end{equation}
Since the nonzero entries can be randomly distributed in the support,  assuming each entry has an initial probability of 0.5 to be nonzero is rational without loss of generality.

Phase 3 - Main iterations. The main process is composed of multiple iterations, and terminates until the halting condition of the algorithm is met. The main process includes the following steps:

1) Candidate supports generation (Line 4):  $N_c$ \emph{candidate supports} $\{ {\Pi _j^{(k)}} \}_{j = 1}^{{N_c}}$  are generated based on the support distribution ${{\bf{q}}^{(k)}}$. Each candidate support ${\Pi _j^{(k)}}$ is generated in an efficient and simple recursive manner to obtain a $K$-sparse support. Let $\pi_l$ denote the current temporary support in the recursive generation process, where the initial temporary support $\pi_0 = \{0, 1, \cdots, N-1\}$. Then, based on the current temporary support $\pi_l$ and its corresponding probability $\{q^{(k)}_n\}_{n \in \pi_l}$ derived from the current support distribution ${{\bf{q}}^{(k)}}$, a more sparse temporary  support $\pi_{l+1}$ can be generated by a Bernoulli trial on each entry $n \in \pi_l$ as
\begin{equation}
\pi_{l+1} = \{n | n \in \pi_l, \textup{and}~ f^{(\pi_l)}_{n} = 1\},
\end{equation}
where the $\{0, 1\}$-valued parameter  $f^{(\pi_l)}_{n} $ is the outcome of the Bernoulli trial on entry $n \in \pi_l$ with Bernoulli probability $q^{(k)}_n$. Afterwards, $l \leftarrow l+1$ and keep doing this until $| \pi_l | \le K$, and then the candidate support is set as ${\Pi _j^{(k)}} = \pi_l$.

2) Computing NBI and residue (Lines 5-6): the estimated NBI vectors $\{ \Delta{\hat{\bf{e}}}_{\textup{B},j}^{(k)} \}_{j = 1}^{{N_c}}$ corresponding to the candidate supports are calculated based on the least squares principle implemented on the candidate supports $\{ {\Pi _j^{(k)}} \}_{j = 1}^{{N_c}}$, and the corresponding residue error norms $\{r_j^{(k)}\}_{j = 1}^{{N_c}}$ are calculated by~\eqref{residue_norm} using the estimated NBI vectors.

3) Favorable supports selection (Lines 7-8): the candidate supports are sorted by the residue error norms in the ascending order in order to pick out the best $N_f$ candidate supports with smallest estimation error, which is closest to the real NBI  support and regarded as the favorable supports $\{ {\Pi_{[j]}^{(k)}} \}_{j = 1}^{{N_f}}$.    The implicit probability distribution implied by the favorable supports is the training target of the current support distribution ${{\bf{q}}^{(k)}}$, which is gradually driven towards the ground-truth distribution by iteratively minimizing the CE between them. \color{black}

4) Learning support distribution by minimizing CE (Line 9):
The CE is utilized as the loss function $\mathcal{L}({\Pi _{[j]}^{(k)}};{{\bf{q}}^{(k)}})$ in the perspective of machine learning theory, which is given by
\begin{equation}\label{loss_CE}
\mathcal{L}({\Pi _{[j]}^{(k)}};{{\bf{q}}^{(k)}}) = { - \frac{1}{{{N_f}}}\sum\limits_{j = 1}^{{N_f}} {\ln \Pr \left( {\left. {\Pi _{[j]}^{(k)}} \right|{{\bf{q}}^{(k)}}} \right)} },
\end{equation}
where $\{-\ln \Pr ( { {\Pi _{[j]}^{(k)}} |{{\bf{q}}^{(k)}}} )\}$ is the negative logarithm likelihood (NLL) of the favorable support ${\Pi _{[j]}^{(k)}}$ conditioned on the current probability distribution ${{\bf{q}}^{(k)}}$. By minimizing the loss function in \eqref{loss_CE}, the current support distribution ${\bf{q}}^{(k)}$ is updated to ${\bf{q}}^{(k+1)}$, which is given by
\begin{equation}\label{q_CEM}
{{\bf{q}}^{(k + 1)}} = \mathop {\arg \min }\limits_{{{\bf{q}}^{(k)}}} \left\{ { - \frac{1}{{{N_f}}}\sum\limits_{j = 1}^{{N_f}} {\ln \Pr \left( {\left. {\Pi _{[j]}^{(k)}} \right|{{\bf{q}}^{(k)}}} \right)} } \right\},
\end{equation}
 Let a $\{0, 1\}$-valued length-$N$ vector ${\bf f}_{[j]}$ denote the favorable support ${\Pi _{[j]}^{(k)}}$, where its $n$-th entry  ${f_{[j],n}} = {({{\bf{f}}_{[j]}})_n}$ satisfies
\begin{equation}\label{vec_f_j}
{f_{[j],n}} = \left\{ \begin{array}{l}
1, ~~n \in \Pi _{[j]}^{(k)}\\
0,~~n \notin \Pi _{[j]}^{(k)}
\end{array} \right.
\end{equation}
Then the conditional probability $\Pr ( { {\Pi _{[j]}^{(k)}} |{{\bf{q}}^{(k)}}})$ in the CE in~\eqref{q_CEM} is given by
\begin{equation}
\Pr ( { {\Pi _{[j]}^{(k)}} |{{\bf{q}}^{(k)}}}) = \Pr ( { {\bf f}_{[j]} |{{\bf{q}}^{(k)}}}),
\end{equation}
where ${f_{[j],n}}$ is a Bernoulli random variable given by
\begin{equation}
\Pr ({f_{[j],n}} = 1) = q^{(k)}_n, ~ \Pr ({f_{[j],n}} = 0) = 1- q^{(k)}_n.
\end{equation}
Thus, one can derive that
\begin{equation}\label{pr_fj_qk}
 \Pr ( { {\bf f}_{[j]} |{{\bf{q}}^{(k)}}}) = \prod\limits_{n = 0}^{N - 1} {{{\left( {q_n^{(k)}} \right)}^{{f_{[j],n}}}}{{\left( {1 - q_n^{(k)}} \right)}^{1 - {f_{[j],n}}}}}.
\end{equation}
By substituting~\eqref{pr_fj_qk} into \eqref{q_CEM}, the first derivative of the CE with respect to $q^{(k)}_n$ can be derived as
\begin{align}
& \frac{\partial }{{\partial q_n^{(k)}}}\left\{ { - \frac{1}{{{N_f}}}\sum\limits_{j = 1}^{{N_f}} {\ln \Pr \left( {\left. {\Pi _{[j]}^{(k)}} \right|{{\bf{q}}^{(k)}}} \right)} } \right\} \nonumber \\
= & \frac{\partial }{{\partial q_n^{(k)}}}\left\{ { - \frac{1}{{{N_f}}}\sum\limits_{j = 1}^{{N_f}} {\left[ {{f_{[j],n}}\ln q_n^{(k)} + (1 - {f_{[j],n}})\ln (1 - q_n^{(k)})} \right]} } \right\} \nonumber \\
= & - \frac{1}{{{N_f}}}\sum\limits_{j = 1}^{{N_f}} {\left[ {\frac{{{f_{[j],n}}}}{{q_n^{(k)}}} - \frac{{1 - {f_{[j],n}}}}{{1 - q_n^{(k)}}}} \right]}. \label{fisrt_derivative}
\end{align}
To minimize the CE, the first derivative \eqref{fisrt_derivative} is set to zero, so the updated support distribution ${{\bf{q}}^{(k+1)}}$ can be learnt by
\begin{equation}\label{update_q}
q_n^{(k + 1)} = \frac{1}{{{N_f}}}\sum\limits_{j = 1}^{{N_f}} {{f_{[j],n}}}, ~n = 0, 1, \cdots, N-1.
\end{equation}

5) Iteration switching (Line 10-11): if the halting condition is satisfied when ${r_{[1]}^{(k-1)}} \le \epsilon $ or $k > I_{\textup m}$, the algorithm ends. Otherwise, the algorithm goes into the next iteration.
\color{black}

Phase 4 - Output. The output of the algorithm includes the learnt support probability distribution ${\hat{\bf{q}}} = {{\bf{q}}^{(k)}}$, the recovered NBI support ${\hat{\Pi}} = {{ \Pi}^{(k-1)}_{[1]}}$, and the recovered sparse NBI vector $\Delta{\hat{\bf{e}}}_{\rm B} =  \Delta{\tilde{\bf{e}}}_{\rm B,[1]}^{(k-1)}$, which obtains the solution of the sparse combinatorial optimization problem in~\eqref{sparse_comb_opt} as $\Delta{\hat{\bf{e}}}_{\rm B}^{*}= \Delta{\hat{\bf{e}}}_{\rm B}$.

Afterwards, ${{{\bf{\tilde e}}}_{{\rm{B}}}}$ can be calculated by~\eqref{de_Bi} and the NBI ${{{\bf{\tilde e}}}_{{\rm{B}}\textup{X}}}$ associated with the OFDM block can be calculated through \eqref{e_BXi}. Then, the NBI can be directly eliminated from the information data in the frequency domain just by subtracting ${{{\bf{\tilde e}}}_{{\rm{B}}\textup{X}}}$ from the received frequency-domain OFDM sub-carriers ${\bf{X}}$, which is given by
\begin{equation}
\label{final_cancel}
{\bf{X}}^{0} = {\bf{X}} - {{{\bf{\tilde e}}}_{{\rm{B}}\textup{X}}},
\end{equation}
\noindent where ${\bf{X}}$ is the DFT of the received OFDM block ${\bf x}_i$ as illustrated in Fig.~\ref{BS_TDM}, while ${\bf{X}}^{0}$ is the frequency-domain OFDM data block free from the   NB-IoT interference. Thus, the NBI-free OFDM data block can be then used for information demapping and decoding.

\subsection{Enhanced Sparse Machine Learning Based Algorithm: Regularized SCEM}
In the proposed SCEM algorithm where the CE plays the important role of loss function, each NLL corresponding to each favorable support ${\Pi _{[j]}^{(k)}}$ has an average contribution to the CE given in \eqref{q_CEM}, so the favorable supports with different residue error norms contribute the same to the loss function. In fact, different supports should reflect different contributions on the loss function so as to encourage the algorithm to learn the support with less error. Out of this insight, an enhanced sparse learning algorithm of RSCEM is proposed, in which the loss function in~\eqref{loss_CE} is regularized by multiplying with the weighting parameter $\lambda_{[j]}$  to generate the regularized loss function $\mathcal{L}_{\textup{reg}}({\Pi _{[j]}^{(k)}};{{\bf{q}}^{(k)}})$ given by
\begin{equation}\label{regularized_loss}
\mathcal{L}_{\textup{reg}}({\Pi _{[j]}^{(k)}};{{\bf{q}}^{(k)}}) = { - \frac{1}{{{N_f}}}\sum\limits_{j = 1}^{{N_f}} {\lambda _{[j]}} {\ln \Pr \left( {\left. {\Pi _{[j]}^{(k)}} \right|{{\bf{q}}^{(k)}}} \right)} },
\end{equation}
where the regularization weighting parameter $\lambda_{[j]}$ is given by
\begin{equation}\label{regularized_parameter}
{\lambda _{[j]}} = \frac{{{{\overline r }^{(k)}}}}{{r_{[j]}^{(k)}}},~ j = 1, 2, \cdots, N_f,
\end{equation}
where ${{{\overline r }^{(k)}}}$ is the average residue error norm over the favorable supports given by
\begin{equation}
{\overline r ^{(k)}} = \frac{1}{{{N_f}}}\sum\nolimits_{j = 1}^{{N_f}} {r_{[j]}^{(k)}}.
\end{equation}

Note that a smaller residue error norm ${{r_{[j]}^{(k)}}}$ leads to a larger weighting parameter ${\lambda _{[j]}}$ in \eqref{regularized_parameter}. Hence, the NLL corresponding to a more accurate support will have a larger contribution to the regularized loss function in \eqref{regularized_loss}, which will drive the support distribution ${{\bf{q}}^{(k)}}$ to converge to the ground-truth support more accurately and more efficiently. The pseudo-code of RSCEM is thus similar to that of SCEM given in \textbf{Algorithm~\ref{Algo1}} except for the procedure of minimizing the loss function in Line 9, where the regularized loss function is now adopted to update the distribution as given by
\begin{equation}\label{regularized_CEM}
{{\bf{q}}^{(k + 1)}} = \mathop {\arg \min }\limits_{{{\bf{q}}^{(k)}}} { - \frac{1}{{{N_f}}}\sum\limits_{j = 1}^{{N_f}}{\lambda _{[j]}} {\ln \Pr \left( {\left. {\Pi _{[j]}^{(k)}} \right|{{\bf{q}}^{(k)}}} \right)} }.
\end{equation}

 To calculate the minimum regularized loss function in \eqref{regularized_CEM}, the same notation as in the previous sub-section, i.e. the Bernoulli vector ${\bf f}_{[j]}$ in \eqref{vec_f_j} denoting the favorable support ${\Pi _{[j]}^{(k)}}$, is inherited. Through similar deduction from \eqref{vec_f_j} to \eqref{pr_fj_qk}, and substituting \eqref{pr_fj_qk} into \eqref{regularized_CEM}, the first derivative of the regularized loss function with respect to $q^{(k)}_n$ can be obtained, represented as \color{black}
\small
\begin{align}
& \frac{\partial }{{\partial q_n^{(k)}}}\left\{ { - \frac{1}{{{N_f}}}\sum\limits_{j = 1}^{{N_f}} {\lambda _{[j]}} {\ln \Pr \left( {\left. {\Pi _{[j]}^{(k)}} \right|{{\bf{q}}^{(k)}}} \right)} } \right\} \nonumber \\
= & \frac{\partial }{{\partial q_n^{(k)}}}\left\{ { - \frac{1}{{{N_f}}}\sum\limits_{j = 1}^{{N_f}} {\lambda _{[j]}} {\left[ {{f_{[j],n}}\ln q_n^{(k)} + (1 - {f_{[j],n}})\ln (1 - q_n^{(k)})} \right]} } \right\} \nonumber \\
= & - \frac{1}{{{N_f}}}\sum\limits_{j = 1}^{{N_f}} {\lambda _{[j]}} {\left[ {\frac{{{f_{[j],n}}}}{{q_n^{(k)}}} - \frac{{1 - {f_{[j],n}}}}{{1 - q_n^{(k)}}}} \right]}. \label{reg_fisrt_derivative}
\end{align}
\normalsize

Setting the first derivative of the regularized loss function given in~\eqref{reg_fisrt_derivative} to zero, the regularized loss function can be minimized,  yielding the updated support probability distribution ${{\bf{q}}^{(k+1)}}$ given by
\begin{equation}\label{reg_update_q}
q_n^{(k + 1)} = \frac{{\sum\limits_{j = 1}^{{N_f}} {{\lambda _{[j]}}{f_{[j],n}}} }}{{\sum\limits_{j = 1}^{{N_f}} {{\lambda _{[j]}}} }}, ~n = 0, 1, \cdots, N-1.
\end{equation}

Comparing \eqref{reg_update_q}  with \eqref{update_q}, it can be observed that, for the algorithm of SCEM, all the entries $\{f_{[j],n}\}_{j=1}^{N_f}$ have the same contribution to the updating of $q_n^{(k+1)}$ in \eqref{update_q}, so the different accuracy among favorable supports are not taken into consideration. On the other hand, for the enhanced RSCEM algorithm, a more accurate support ${\Pi _{[j]}^{(k)}}$ will impose a larger weighting parameter ${\lambda _{[j]}}$ on and have a larger contribution to the updating of $q_n^{(k+1)}$ as implied by \eqref{reg_update_q}. In fact, \eqref{update_q} can be regarded as a special case of \eqref{reg_update_q} when ${\lambda _{[j]}} = 1, ~ j = 1, 2, \cdots, N_f$. Consequently, it can be derived that the enhanced RSCEM algorithm will learn the ground-truth support distribution more accurately and  more efficiently than  SCEM, which is also validated in the simulation results in the next section.

\subsection{Extension to MIMO: Simultaneous Multi-Antenna NBI Recovery Algorithm}
The proposed method can be extended to MIMO systems  to further improve the estimation accuracy by exploiting the spatial correlation of the NBI. Due to the spatial correlation, the received NBI signals at different receive antennas in the MIMO system share the same support, i.e., the locations of nonzero entries are the same, although their amplitudes might be different~\cite{S-Liu6}. One can make use of the spatial correlation in the iterations of the proposed sparse machine learning algorithm to simultaneously recover the NBI signals contaminating multiple receive antennas.

\begin{algorithm}                      

\caption{(\emph{S-SCEM}): Simultaneous Sparse Cross-Entropy Minimization for NBI Recovery in MIMO System}              
\label{Algo2}                           
\begin{algorithmic}[1]                    
\REQUIRE~~\\
1) Measurement vectors $\{ \Delta {\bf p}_{t} \}_{t=1}^{N_{\textup{R}}}$ at $N_{\textup{R}}$ receive antennas\\
2) Observation matrix ${\bf{\Psi }} = {\bf{S}}_{G,N}{{\bf{F}}_N}$\\
3) Threshold for residue error norm  $\epsilon$\\
4) Parameters $N_c$, $N_f$, $I_{\textup m}$\\

\textbf{\textit{Initialization:}}
\STATE ${{\bf{q}}^{(0)}} \leftarrow \frac{1}{2} \cdot {\overrightarrow {\bf{1}} _{N \times 1}}$
\STATE $k \leftarrow 0$ \\

\textbf{\textit{Iterations:}}
\REPEAT
	\STATE Randomly generate $N_c$ \emph{candidate supports} $\{ {\Pi _j^{(k)}} \}_{j = 1}^{{N_c}}$ based on the current support distribution ${\bf{q}}^{(k)}$
	 s.t. $| {\Pi _j^{(k)}} | \le {K}, j = 1, \cdots N_c$\\
	\STATE For each candidate support ${\Pi _j^{(k)}}, j = 1, \cdots N_c$, compute $N_{\textup{R}}$  NBI vectors $\{ \Delta{\tilde{\bf{e}}}_{\textup{B},j(t)}^{(k)} \}_{t = 1}^{N_{\textup{R}}}$, s.t. \\ ${\left. \Delta{\tilde{\bf{e}}}_{\textup{B},j(t)}^{(k)} \right|_{{{\Pi^{(k)}_j}}}} \leftarrow {\bf{\Psi }}_{{{\Pi ^{(k)}_j}}}^\dagger {\Delta {\bf p}_{t}}, ~~ {\left. \Delta{\tilde{\bf{e}}}_{\textup{B},j(t)}^{(k)} \right|_{{\Pi ^{(k)c}_j}}}  \leftarrow {\bf{0}}$\\
	\STATE For each candidate support ${\Pi _j^{(k)}}, j=1,\cdots N_c$, calculate $N_{\textup{R}}$ residue error norms:\\ $\{ {r_{j(t)}^{(k)}} = {\| {{\Delta {\bf p}_{t}} - {\bf{\Psi }} \Delta{\tilde{\bf{e}}}_{\textup{B},j(t)}^{(k)}} \|_2} \}_{t = 1}^{N_{\textup{R}}}$\\
	\STATE Calculate the weighted average residue error norms $\{ \bar{r}_j^{(k)} \}_{j = 1}^{{N_c}}$ by: \\ $\bar{r}_j^{(k)} = \sum_{t = 1}^{N_{\textup{R}}} {{\beta_t} \cdot {r_{j(t)}^{(k)}}}$, with $\beta_t$ given by \eqref{beta}
	
	\STATE Sort the weighted average residue error norms in the ascending order as\\ $\bar{r}_{[1]}^{(k)} \le \bar{r}_{[2]}^{(k)} \le  \cdots  \le \bar{r}_{[{N_c}]}^{(k)}$\\
	\STATE Select the $N_f$ smallest weighted average residue error norms $\{ {\bar{r}_{[j]}^{(k)}} \}_{j = 1}^{{N_f}}$, and set $\{ {\Pi_{[j]}^{(k)}} \}_{j = 1}^{{N_f}}$ as the \emph{favorable supports}\\
	\STATE Update the probability distribution of NBI support to ${\bf{q}}^{(k+1)}$ by minimizing the CE based on~\eqref{q_CEM}\\
	\STATE $k \leftarrow k + 1$\\
\UNTIL {${\bar{r}_{[1]}^{(k-1)}} \le \epsilon $ or $k > I_{\textup m}$}  ~~~~~~~  (halting condition)\\

\ENSURE~~\\
1) Learnt support probability distribution ${\hat{\bf{q}}} = {{\bf{q}}^{(k)}}$\\
2) Estimated common sparse NBI support ${\hat{\Pi}} = {{ \Pi}^{(k-1)}_{[1]}}$\\
3) Estimated  $N_{\textup{R}}$ NBI vectors $\{ \Delta{\hat{\bf{e}}}_{\textup{B}(t)} =  \Delta{\tilde{\bf{e}}}_{{\rm B,[1]}(t)}^{(k-1)}\}_{t = 1}^{N_{\textup{R}}}$  \\
\end{algorithmic}
\end{algorithm}

Specifically, the SCEM algorithm in  \textbf{Algorithm~\ref{Algo1}} can be extended  to the MIMO system to formulate the algorithm called  simultaneous SCEM (S-SCEM), whose details are presented in \textbf{Algorithm~\ref{Algo2}}.
 The input includes the measurement vectors  at $N_{\textup{R}}$ receive antennas, denoted by $\{ \Delta {\bf p}_{t} \}_{t=1}^{N_{\textup{R}}}$, obtained by the temporal differential measuring operations on the $N_{\textup{R}}$ receive antennas as in~\eqref{BSBL} for the single antenna case. Other input parameters and the initialization process are similar to those of SCEM. For the $k$-th iteration in the repeated loop, firstly, the $N_c$ candidate supports $\{ {\Pi _j^{(k)}} \}_{j = 1}^{{N_c}}$ are generated in the same way as that of SCEM. Then, for each candidate support ${\Pi _j^{(k)}}$, the $N_{\textup{R}}$ NBI vectors $\{ \Delta{\tilde{\bf{e}}}_{\textup{B},j(t)}^{(k)} \}_{t = 1}^{N_{\textup{R}}}$ and residue error norms $\{ {r_{j(t)}^{(k)}} \}_{t = 1}^{N_{\textup{R}}}$ corresponding to the $N_{\textup{R}}$ receive antennas are calculated. Note that the method of sorting the residue error norms is different from SCEM in the MIMO case. In order to take the contributions of all the receive antennas into account, the residue error norms over the $N_{\textup{R}}$ receive antennas can be summed up for each candidate support ${\Pi _j^{(k)}}$ before sorting them. An alternative approach is weighted averaging, where the weights $\{ \beta_t \}_{t=1}^{N_{\textup{R}}}$ are proportional to the signal-to-noise ratio (SNR) given by
\begin{equation}\label{beta}
\beta_t  = \frac{\rho_t}{\sum_{r = 1}^{N_{\textup{R}}} {\rho_r}}, t = 1, \cdots, N_{\textup{R}},
\end{equation}
\noindent where $\rho_t$ is the SNR at the $t$-th receive antenna and can be estimated through the pilot power versus the noise floor level. Thus the weighted average $\bar{r}_j^{(k)}$ of the residue error norms for candidate support ${\Pi _j^{(k)}}$ is given by
\begin{equation}\label{weighted_average}
\bar{r}_j^{(k)} = \sum_{t = 1}^{N_{\textup{R}}} {{\beta_t} \cdot {r_{j(t)}^{(k)}}}.
\end{equation}
In \textbf{Algorithm~\ref{Algo2}}, the weighted average is adopted to determine the average residue error and sort the candidate supports. After summing or weighted averaging, the information of the residue error norms from all the receive antennas can be exploited to sort the candidate supports, and generate $N_f$ favorable supports $\{ {\Pi_{[j]}^{(k)}} \}_{j = 1}^{{N_f}}$ by picking out the best $N_f$ ones. Afterwards, the current probability distribution of the support is similarly updated by minimizing the CE according to~\eqref{q_CEM}. The halting condition is similar to that of SCEM when weighted average is adopted, while the threshold should be $N_{\textup{R}} \epsilon$ when summing is adopted before sorting. Finally, the output of S-SCEM is the estimated common NBI support ${\hat{\Pi}}$ and the estimated $N_{\textup{R}}$ NBI vectors $\{ \Delta{\hat{\bf{e}}}_{\textup{B}(t)} \}_{t = 1}^{N_{\textup{R}}}$ that can be utilized to eliminate all the NBI signals at the $N_{\textup{R}}$ receive antennas, respectively.
\color{black}

\section{Performance Evaluation and Simulation Results}\label{Sec-V}

\subsection{Computational Complexity Analysis}
The computational complexity of the proposed algorithms are theoretically and numerically analyzed and compared as follows.

For the proposed algorithms, considering the complexity of each iteration of SCEM in \textbf{Algorithm 1}: Line 4 (generating $N_c$ candidate supports) - ${\mathcal O}\left( N_c \right)$; Line 5 (calculating $N_c$ NBI vectors) - ${\mathcal O}\left( N_cGK^2 \right)$; Line 6 (calculating $N_c$ residue error norms) - ${\mathcal O}\left( N_cGK \right)$; Lines 7 - 8  (sorting $N_c$ residue error norms and selecting $N_f$ smallest ones) - ${\mathcal O}\left( N_c \log{N_c} \right)$; Line 9 (updating the NBI support distribution) - ${\mathcal O}\left( NN_f \right)$. Therefore, summing them together, the total complexity of each iteration of SCEM is  ${\mathcal O}\left( N_cGK^2+NN_f \right)$. Similarly, since RSCEM only involves the calculation of $N_f$ weighting parameters in~\eqref{regularized_parameter} with the complexity of ${\mathcal O}\left(N_f \right)$, one can derive that the total complexity of each iteration of RSCEM is also ${\mathcal O}\left( N_cGK^2+NN_f \right)$.
 Then, considering the maximum iteration number $I_{\textup m}$,  the total complexity of the SCEM and RSCEM algorithms are ${\mathcal O}\left( I_{\textup m}(N_cGK^2+NN_f) \right)$.  The complexity of the S-SCEM algorithm for MIMO systems can be similarly derived as ${\mathcal O}\left( I_{\textup m}(N_{\textup{R}}N_cGK^2+NN_f) \right)$. \color{black} Compared with the existing block SBL-based algorithm which costs a complexity of ${\mathcal O}\left( u^2N^2G \right)$, where $u$ is another parameter related with the NBI block distribution~\cite{SLiu17TCOM}, the proposed algorithms can recover the NBI with an acceptable and comparable complexity.

From the above analysis, it can be noted that apart from the sparsity level $K$ of the NBI, the complexity performance of the proposed algorithm is mainly dependent on the choice of the parameters $N_c$, $N_f$, and $I_{\textup m}$. \color{black}
	 If a larger value of $N_c$ and $N_f$ is chosen, the computational complexity will be linearly increased. On the other hand, since more candidate supports are generated from the distribution, and more favorable supports with smallest estimation error are selected to calculate the CE for the learning and training process, each iteration will be further approaching the ground-truth distribution, which makes the algorithm to converge more rapidly.  Thus, the required total maximum iteration number $I_{\textup m}$ can be reduced in order to reach the halting condition of estimation error given by ${r_{[1]}^{(k-1)}} \le \epsilon $ in \textbf{Algorithm~\ref{Algo1}}.

	Therefore, there is a tradeoff between the computational complexity of each iteration and the total number of iterations $I_{\textup m}$.  \color{black} If the learning agent (e.g. LTE base station) has more available computing resource and hopes to deal with a real-time NBI estimation and cancellation, then choosing a larger $N_c$ and $N_f$ is more suitable. Otherwise if computing resource is the bottleneck but longer delay is tolerable such as for some cost-effective terminal, $N_c$ and $N_f$ can be set smaller.   Empirically, $N_c$=70, $N_f$=15 and $I_{\textup m}=15$ are set in the following simulations in this work, which leads to a moderate computational complexity and convergence rate.   \color{black}

Furthermore, the performance of the parameters influencing the computational complexity of the proposed algorithm is evaluated through numerical analysis as follows.

  First, the parameters of the number of candidate and favorable supports $N_c$, $N_f$, and the maximum iteration number $I_{\textup m}$ are investigated in order to reach successful NBI estimation.  \color{black}  A successful NBI estimation is recognized if the support is recovered correctly and the mean square error (MSE) of estimation is smaller than ${10^{- 3}}$. The numerical analysis result for the parameters  is   listed in Table~\ref{Tab:parameter_choice}, where the NBI signal dimension $N$ is assumed to be fixed at 600. {As specified in the 3GPP LTE standards~\cite{LTE1}, the active data OFDM sub-carrier number is set as $N=600$ (when the number of resource block is 50), which is the signal dimension of the NBI.} It can be observed that the numerical analysis is consistent with the theoretical analysis above.   For example, when considering $K=26$, if a larger value of $N_c=98$ and $N_f=26$ is chosen, $I_{\textup m}$ can be reduced to 15 compared with 23.  \color{black} In this way, the convergence rate is faster due to fewer iterations, but the cost is that each iteration is more computationally complicated.    Besides, when we investigate all over the three sparsity levels $K=13, 26, 39$, it can be observed that when $N_c=70$ and $N_f=15$ are fixed, then $I_{\textup m}$ should be set larger for a larger $K$ to allow more iterations and longer algorithm delay.

However, the overall computational complexity ${\mathcal O}\left( I_{\textup m}(N_cGK^2+NN_f) \right)$ has similar order for different parameter choices with the same sparsity level because it is observed from Table~\ref{Tab:parameter_choice} that $I_{\textup m}$ changes in the opposite direction to $N_c$ and $N_f$.  \color{black} Moreover, it can be noted that with the rapid linear increase of $K$, the parameters required to reach successful NBI estimation is not increasing as fast and the overall complexity is kept at an acceptable level in different conditions. This validates the advantage and efficiency of the proposed sparse machine learning algorithm in training and approximating the ground-truth distribution rapidly compared with blind random exploration.

\begin{table}[t]

\caption{Parameters of Proposed Algorithm  to Reach Successful NBI Recovery with Different Sparsity Level $K$.}
\renewcommand{\arraystretch}{1.2}
\vspace{-15pt}
\small
\label{Tab:parameter_choice}
\begin{center}
\begin{tabular}{*{1}{C{0.8cm}}|*{1}{C{0.8cm}}|*{1}{C{0.8cm}}|*{1}{C{0.8cm}}|*{1}{C{0.8cm}}|*{1}{C{0.8cm}}|*{1}{C{0.8cm}}}
 \hline\hline
   $K$ & \multicolumn{2}{c|}{13} & \multicolumn{2}{c|}{26} & \multicolumn{2}{c}{39}   \\
  \hline
   $I_{\textup m}$ & 15 &  20    & 15  & 23 &  15  &  31\\
     \hline
   $N_c$    & 70 &   62  &  98  & 70 & 143 &  70\\
        \hline
   $N_f$    & 15 &   11  &  26  & 15   & 32  &  15\\
  \hline \hline
\end{tabular}
\end{center}
\normalsize
\end{table}

Second, the actual number of iterations needed upon successful estimation (MSE $<10^{-3}$)  with respect to sparsity level $K$  is investigated for both state-of-the-art and the proposed algorithms through numerical analysis, as shown in Table~\ref{Tab:Iteration_K}. {The performance of the state-of-the-art CS-based algorithms called sparsity adaptive matching pursuit (SAMP)~\cite{T-Do} and priori aided SAMP (PA-SAMP)~\cite{S-Liu5} is also evaluated for comparison.} It is shown that the actual number of iterations for CS-based algorithms increase approximately linear with sparsity level $K$. On the other hand, the actual number of iterations needed by the proposed algorithms, which reflects the convergence rate, almost keeps invariant, which is consistent with the theoretical analysis. This is because $N_c$ and $N_f$ can be adjusted accordingly with the increase of $K$,  although each iteration will cost more computational complexity.     It is also verified that setting the maximum iteration number  as $I_{\textup m} = 15$ for all the three proposed algorithms is sufficient to  recover the NBI accurately.   \color{black}
Thus the processing delay of the proposed algorithms in the learning agent is properly guaranteed and constrained.
Besides, it is observed from Table~\ref{Tab:Iteration_K} that, the proposed enhanced algorithms of RSCEM and S-SCEM need fewer iterations than SCEM and thus converge faster, which shows that the regularization of the loss function and the exploitation of the spatial correlation can bring significant benefit to the training and learning process.

\begin{table}[h!]
\caption{The Average Number of Actual Iterations of Different NBI Recovery Algorithms w.r.t Sparsity Level $K$.}
\renewcommand{\arraystretch}{1.2}
\vspace{-15pt}
\small
\label{Tab:Iteration_K}
\begin{center}
\begin{tabular}{*{1}{C{0.8cm}}|*{1}{C{1.1cm}}|*{1}{C{1.1cm}}|*{1}{C{1.1cm}}|*{1}{C{1.1cm}}|*{1}{C{1.2cm}}}
 \hline\hline
  sparsity level $K$ & Conv. PA-SAMP & Conv. SAMP &  Prop. SCEM &  Prop. RSCEM & Prop. S-SCEM \\
  \hline
   13 &  10.5    & 12.8  & 12.5 &  10.6  &  8.3\\
     \hline
   26 &   22.4  & 26.1  & 12.4 & 10.5 &  8.5\\
        \hline
   39 &   34.8  & 38.6  & 12.6 & 10.7  &  8.8\\
  \hline \hline
\end{tabular}
\end{center}
\normalsize
\end{table}

\subsection{Simulation Results of NBI Estimation and Elimination}

 The performance of the proposed methods for the estimation and elimination of the NB-IoT interference to the LTE system is evaluated by extensive simulations. \color{black}  The proposed algorithms of SCEM and RSCEM are simulated in the single-input single-output (SISO) system, while the proposed S-SCEM algorithm is simulated in a $4 \times 4$ MIMO system with $N_{\textup{R}} = 4$ using weighted averaging given in~\eqref{weighted_average} as the  criterion of sorting and selecting the favorable supports.

\begin{figure}[t]
\centering
\includegraphics[width=0.4\textwidth]{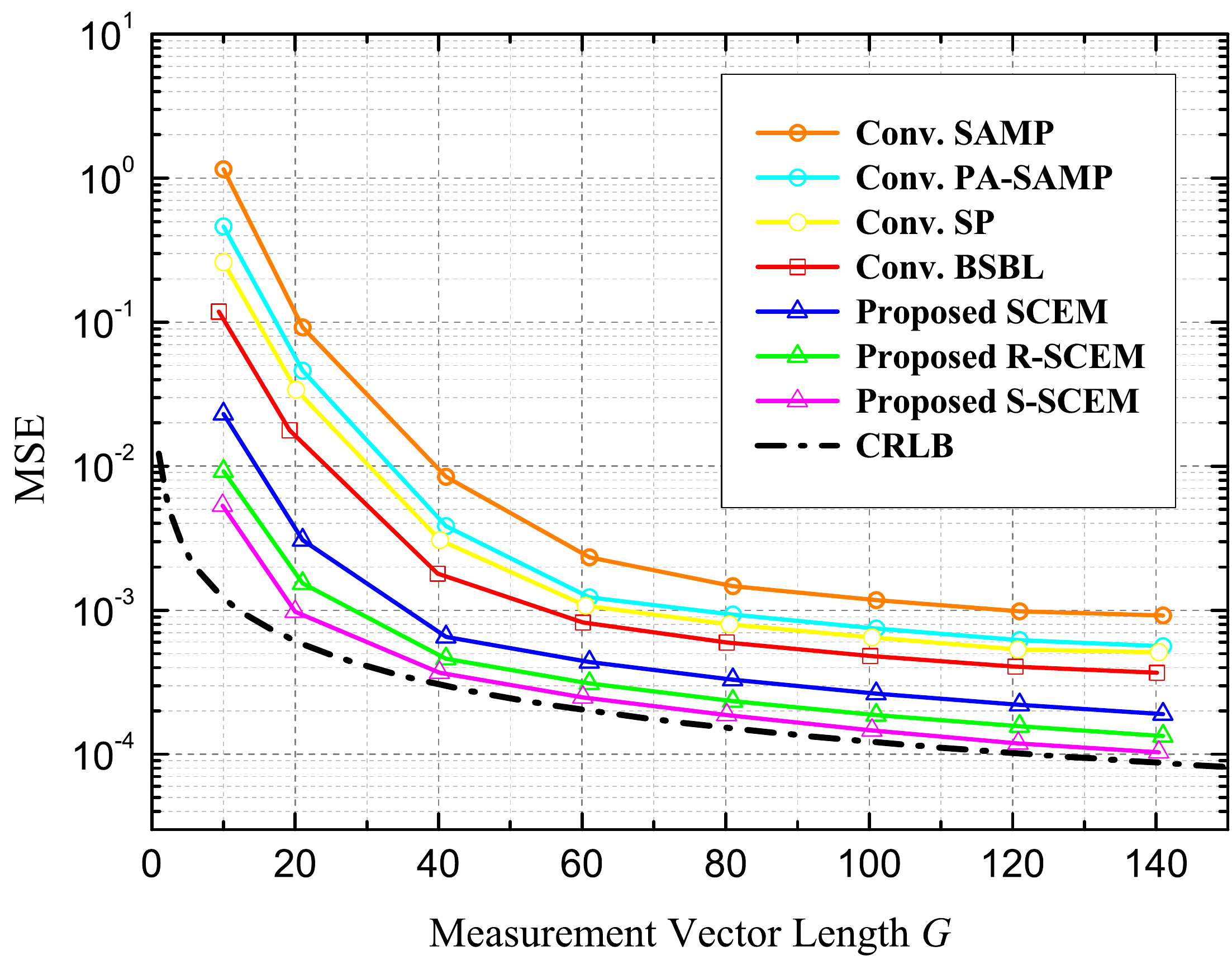}
\caption{MSE performance of the NBI recovery using the proposed and conventional algorithms with respect to the measurement vector length $G$.}
\label{MSE_G}
\end{figure}

{As specified in the 3GPP LTE standards~\cite{LTE1}, the length of each CP is set as $V = 144$ when the active data OFDM sub-carrier number is $N=600$.} The sub-carrier spacing is 15 kHz, so the occupied active data bandwidth is configured as 9.0 MHz~\cite{LTE1}, leading to a CP duration of 4.68 $\mu s$. In this operation mode of LTE, the total number of sub-carriers considering inactive and other ones is 1024, and the total channel bandwidth is 10.0 MHz~\cite{LTE1, LTE2}. The equivalent baseband multipath six-tap channel called the \mbox{ITU-R} Vehicular-A channel model~\cite{V-B}, which is widely used to emulate the wireless mobile channel, is applied, where the UE receiver velocity of 20 km/h is used to present the typical low-speed mobile channel. {The maximum delay spread of the Vehicular-A channel is 2.51 $\mu s$, which is equivalent to the channel length $L = 76$, so the size of the IBI-free region is $G = 68$. In the simulations, the size of the IBI-free region can be pre-determined according to the system configuration of frame length and the maximum channel delay spread of the adopted channel. According to the CS theory in literature, the CS-based methods require the measurement vector length $G$ to be in order of ${\mathcal O}\left(K \log (N) \right)$~\cite{D-Donoho}, which means the size of the IBI-free region is sufficient for effective recovery with overwhelming high probability. Based on the simulation results that will be reported in Fig.~\ref{MSE_G}, setting the measurement vector length $G$ to be 68 is more than sufficient for accurate recovery using the proposed sparse machine learning based algorithms. In realistic implementation, the maximum channel delay spread can also be obtained from the \emph{a priori} knowledge of the channel environment and channel statistics, or from the coarse channel estimation of the path delays using the correlation of training sequences at the receiver~\cite{L-Dai3}.}  The turbo code with code rate of 1/3 as well as the 64QAM modulation as specified in the LTE  standards~\cite{LTE1} are adopted.

As described in Section~\ref{III-B}, each tone interferer of the NBI generated by the NB-IoT signal follows a Gaussian distribution. The FO of the NBI is configured to be \emph{a priori} known as $\alpha = 0.20$ in the simulations, while it can also be effectively estimated at the receiver through the grid search method~\cite{A-Gom2} in realistic implementation. Since each NB-IoT service occupies a bandwidth of 200 kHz according to the NB-IoT specifications~\cite{NB-IoT1}, which is equivalent to 13 sub-carriers in the LTE spectrum, the sparsity level of the NBI is assumed to be $K=13$. To make the NBI model more general, the support $\Pi$ of the NBI is assumed to follow a uniform distribution $\textup{U}\left[0,N-1\right]$ among all the $N$ sub-carriers. Unless otherwise specifically stated, the INR of the NBI is configured as $\gamma = 15$ dB in the simulations.

As described in the previous numerical analysis, the parameters $N_c = 70, N_f = 15$, and $I_{\textup m} = 15$ are proper configuration for the proposed algorithms of SCEM, RSCEM, and S-SCEM. The performance of the state-of-the-art CS-based algorithms including subspace pursuit (SP)~\cite{W-Dai}, SAMP~\cite{T-Do} and PA-SAMP~\cite{S-Liu5}, as well as the block SBL-based algorithm called block SBL (BSBL)~\cite{SLiu17TCOM}, is also evaluated and reported for comparison. The simulation is carried out by Matlab R2017b  on the platform of  Intel Core i7 with frequency of 2.80GHz and RAM of 8.00GB. In the evaluation, the learning agent is the agent operating the proposed sparse machine learning algorithms on a wireless cellular transmission terminal specified by 3GPP LTE standards~\cite{LTE1}, in the presence of the interference from NB-IoT system.
 \color{black}

\begin{figure}[t]
\centering
\includegraphics[width=0.4\textwidth]{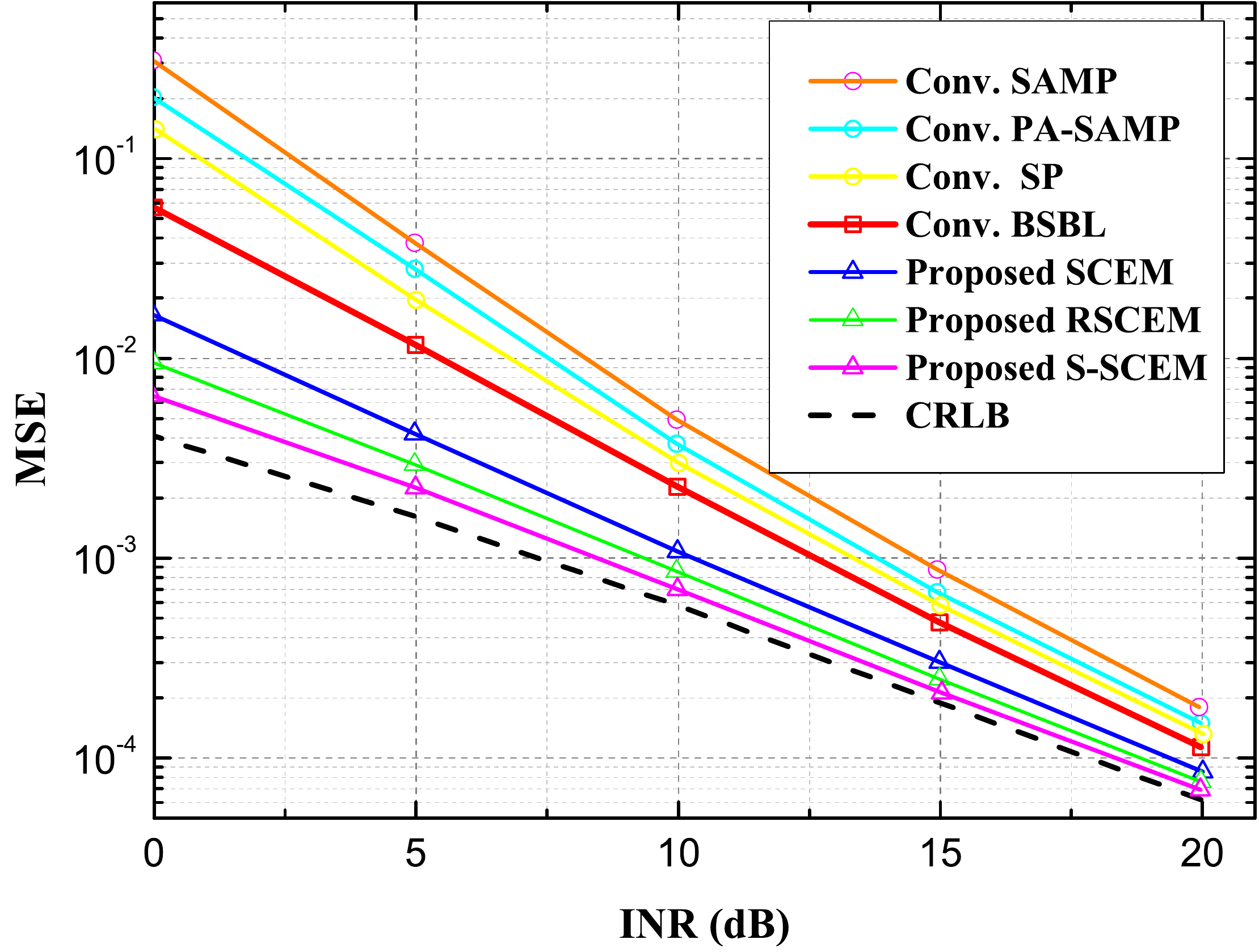}
\caption{MSE performance comparisons of the proposed sparse machine learning based method and the conventional counterparts for NBI recovery in the LTE system under the wireless Vehicular-A channel.}
\label{NBI_MSE}
\end{figure}

{
  The mean square error (MSE) performance of NBI recovery using the proposed methods are shown in Fig.~\ref{NBI_MSE}, with the y-axis being logarithmic.  \color{black}  The performance of the proposed sparse machine learning based methods (SCEM, RSCEM, and S-SCEM with $4 \times 4$ MIMO configuration), the conventional SBL-based algorithm BSBL~\cite{SLiu17TCOM}, and the conventional CS-based methods (PA-SAMP~\cite{S-Liu5} and SAMP~\cite{T-Do}) are depicted. The theoretical Cramer-Rao lower bound (CRLB) calculated by $2\sigma_w^2(K/G)$~\cite{S-Liu5,CRLB-book} is also included as a benchmark. It is noted from Fig.~\ref{NBI_MSE} that the proposed algorithms of S-SCEM, RSCEM, and SCEM achieve a target MSE of ${10^{ - 3}}$ at the INR of 8.6 dB, 9.3 dB, and 10.4 dB, respectively. It is demonstrated that the proposed S-SCEM algorithm has the best performance by exploiting the spatial correlation of the NBI to simultaneously improve the estimation accuracy. It is also shown that the proposed enhanced algorithm RSCEM achieves a further INR gain of about 1.1~dB over the SCEM algorithm, by imposing regularization on the loss function. {Moreover, it can be observed that the proposed approaches outperform the conventional sparse approximation algorithms of BSBL, SP, PA-SAMP, and SAMP by approximately 3.3~dB, 4.0 dB, 4.5 dB, and 5.3 dB, respectively.} It is also noted from Fig.~\ref{NBI_MSE} that the MSE of the proposed algorithms are asymptotically approaching the theoretical CRLB with the increase of the INR, which verifies the high accuracy of the proposed sparse learning method for NBI recovery.
}

Moreover, it is shown in Fig.~\ref{MSE_G} that the MSE of the proposed algorithms decreases fast with the increase of the measurement vector length $G$, i.e., the length of the received IBI-free region ${\bf p}$ in~\eqref{q_i} utilized for NBI measurement as shown in Fig.~\ref{BS_TDM}, whereas the decrease of the MSE of the coventional sparse approximation methods is much slower. {At the MSE of $10^{-3}$, the proposed algorithms of S-SCEM, RSCEM, and SCEM cost only $G=20$, 28, and 35 time-domain samples of the IBI-free region, respectively, whereas the SBL-based and CS-based algorithms cost more than 55 and 65 samples.} Hence, it can be concluded that the introduction of sparse machine learning will greatly reduce the amount of measurement data required for accurate recovery, achieving higher spectral efficiency than conventional counterparts.

\begin{figure}[t]
\centering
\includegraphics[width=0.4\textwidth]{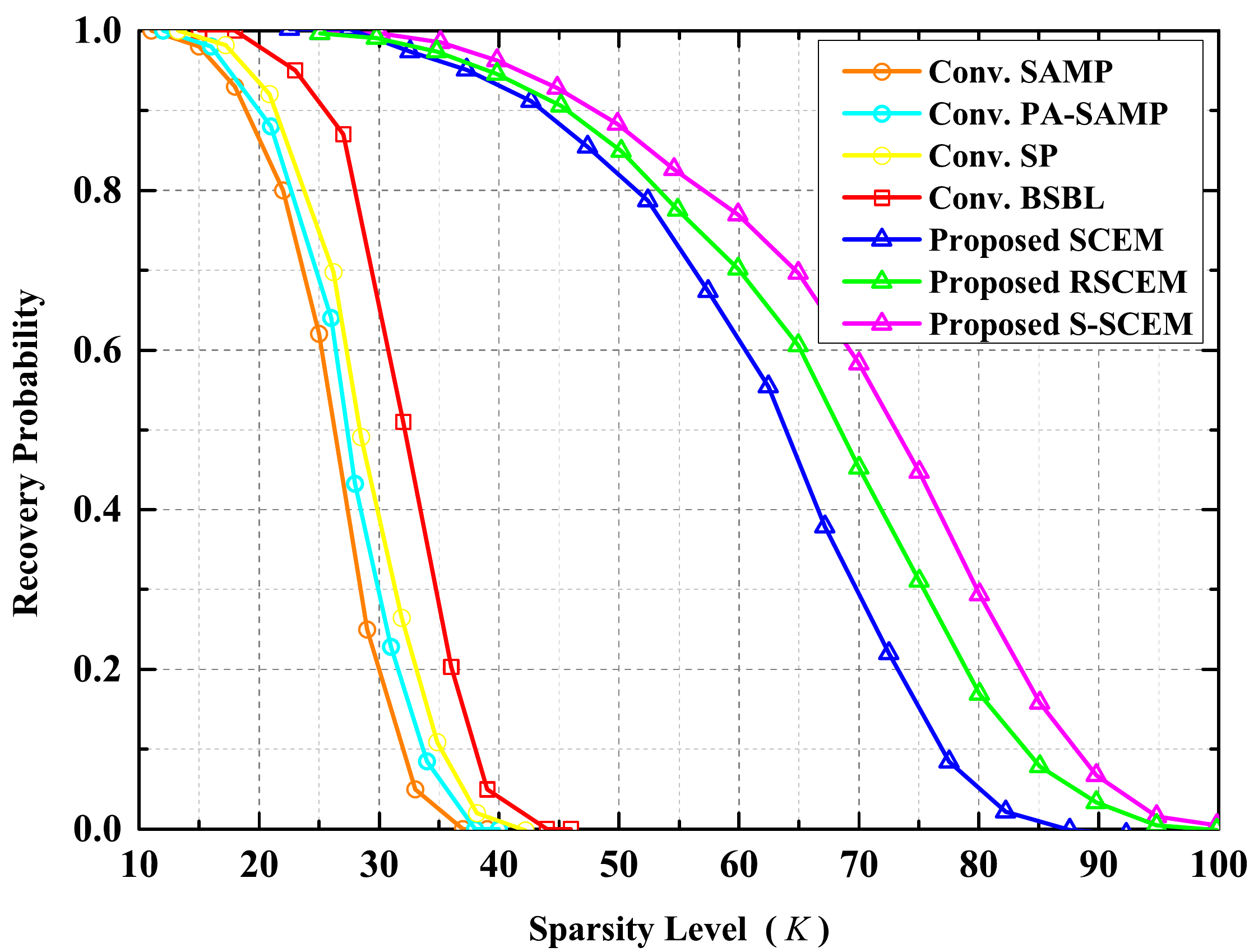}
\caption{Probability of NBI recovery using the proposed and conventional methods in the LTE system under the wireless Vehicular-A channel.}
\label{Recover_Pr_K}
\end{figure}

The recovery probability of the proposed method for NBI recovery versus the sparsity level $K$ under the Vehicular-A channel is depicted in Fig.~\ref{Recover_Pr_K}. The recovery probability is defined as the probability of the effective NBI estimation (i.e. correct support estimation and MSE $<{10^{- 3}}$), which is calculated by the ratio of the number of effective NBI estimations to the total $10^3$ simulations in Fig.~\ref{Recover_Pr_K}. {It is noted that the proposed algorithms reach a successful recovery probability of 0.90 at the sparsity level of $K  = 45$, which significantly outperforms the conventional SBL-based and CS-based algorithms with $K=26$ and $K = 22$, respectively.} It is thus validated that the proposed methods can accurately recover the NBI with much larger sparsity levels using limited measurement data compared with the conventional sparse approximation algorithms. Since each NB-IoT signal occupies 13 sub-carriers in the LTE spectrum, it can be inferred that the proposed method is capable of effectively recovering and eliminating at least 3 ``in-band'' NB-IoT interfering signals in the LTE system.

\begin{figure}[t]
\centering
\includegraphics[width=0.4\textwidth]{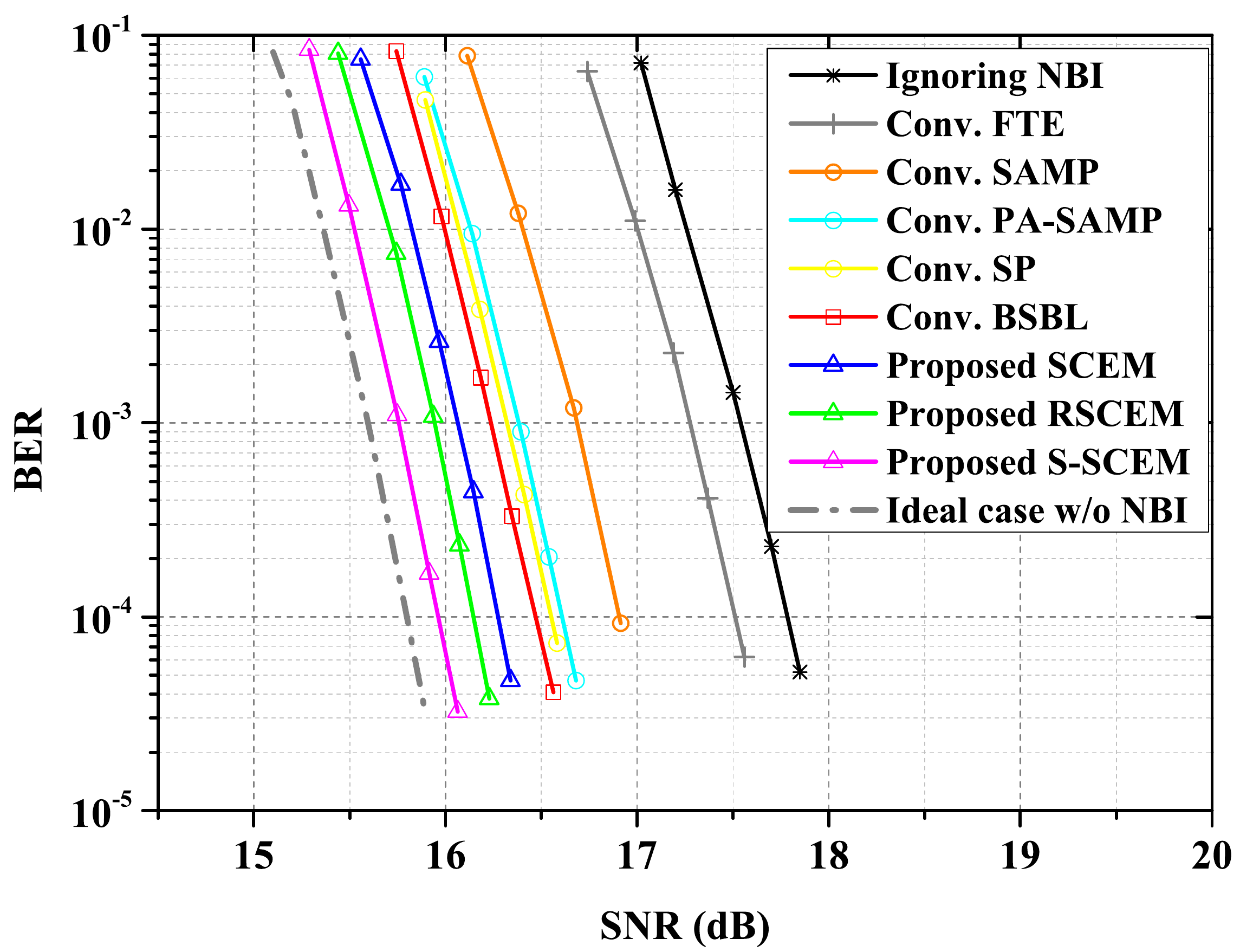}
\caption{BER performance comparison of different NBI mitigation schemes in the LTE system under wireless Vehicular-A channel in the presence of NBI.}
\label{BER}
\end{figure}

{
The bit error rate (BER) performance of the proposed method at the UE receiver in the LTE system under the wireless Vehicular-A channel is illustrated in Fig.~\ref{BER}. Apart from the conventional CS-based and SBL-based algorithms~\cite{T-Do, S-Liu5, SLiu17TCOM}, the BER performance of the conventional FTE method~\cite{S-Kai} is also reported for comparison. The worst case ignoring NBI and the ideal case without NBI are also depicted as benchmarks. {It can be observed that at the target BER of ${10^{ - 4}}$, the proposed sparse machine learning based algorithms  significantly outperform the state-of-the-art SBL-based algorithm, the  existing CS-based  algorithms, the traditional FTE method, and the case ignoring NBI by about 0.5 dB, 0.7 dB, 1.5 dB, and 1.8~dB, respectively.} This implies that the NBI can be more effectively recovered and eliminated in the proposed probabilistic framework of sparse machine learning using the iterative learning algorithms compared with  state-of-the-art counterparts. Furthermore, it is shown in Fig.~\ref{BER} that the gap between the curves of the proposed algorithms and the ideal case without NBI is only about 0.2~dB, validating the accuracy and effectiveness of the proposed schemes for NBI mitigation in the heterogeneous networks composed of NB-IoT and LTE systems.

}

\section{Conclusions}
In this paper, a novel sparse machine learning based probabilistic framework of NBI recovery is formulated  for harmonic coexistence  of NB-IoT and LTE systems. The original non-convex sparse combinatorial optimization problem of NBI recovery is efficiently and accurately solved by the proposed sparse learning algorithm of SCEM, which iteratively learns the probability distribution, i.e. the sparse pattern, of the NBI support by minimizing the loss function of cross-entropy. By imposing regularization on the loss function, the enhanced algorithm of RSCEM is proposed to further improve the convergence rate and accuracy. Furthermore, the spatial correlation of the NBI in multiple receive antennas of the MIMO system is exploited to simultaneously recover the NBI signals more accurately and efficiently. It is verified by theoretical analysis and numerical simulation results that the proposed algorithms outperform state-of-the-art counterparts in spectrum efficiency, estimation accuracy and computational complexity. Using the proposed method, both the UEs and the base stations in LTE systems can be protected from the contamination of NB-IoT interference. Moreover, the proposed scheme can also be widely applied in other wireless heterogeneous networks and broadband systems contaminated by NBI.

\bibliography{IEEEabrv,IoTJ_NBIoT_bibfile}

\end{document}